\documentclass[review]{elsarticle}

\makeatletter
\def\@author#1{\g@addto@macro\elsauthors{\normalsize%
    \def\baselinestretch{1}%
    \upshape\authorsep#1\unskip\textsuperscript{%
      \ifx\@fnmark\@empty\else\unskip\sep\@fnmark\let\sep=,\fi
      \ifx\@corref\@empty\else\unskip\sep\@corref\let\sep=,\fi
      }%
    \def\authorsep{\unskip,\space}%
    \global\let\@fnmark\@empty
    \global\let\@corref\@empty  
    \global\let\sep\@empty}%
    \@eadauthor={#1}
}
\makeatother

\usepackage{hyperref}
\usepackage{lineno}
\modulolinenumbers[5]

\usepackage{graphicx}
\graphicspath{{Images/}}

\usepackage{subfigure}
\usepackage{color}
\usepackage{amsmath}

\journal{}

\biboptions{square,numbers,sort&compress}

\begin{document}

\begin{frontmatter}

\title{An explicit relaxation filtering framework based upon Perona-Malik anisotropic diffusion for shock capturing and subgrid scale modeling of Burgers turbulence}

\author{Romit Maulik \corref{mycorrespondingauthor}}
\cortext[mycorrespondingauthor]{Corresponding author}
\ead{romit.maulik@okstate.edu}
\author{Omer San}
\ead{osan@okstate.edu}
\address{School of Mechanical And Aerospace Engineering, Oklahoma State University, Stillwater, Oklahoma 74078, USA}

\begin{abstract}

In this paper, we introduce a relaxation filtering closure approach to account for subgrid scale effects in explicitly filtered large eddy simulations using the concept of anisotropic diffusion. We utilize the Perona-Malik diffusion model and demonstrate its shock capturing ability and spectral performance for solving the Burgers turbulence problem, which is a simplified prototype for more realistic turbulent flows showing the same quadratic nonlinearity. Our numerical assessments present the behavior of various diffusivity functions in conjunction with a detailed sensitivity analysis with respect to the free modeling parameters. In comparison to direct numerical simulation (DNS) and under-resolved DNS results, we find that the proposed closure model is efficient in the prevention of energy accumulation at grid cut-off and is also adept at preventing any possible spurious numerical oscillations due to shock formation under the optimal parameter choices. In contrast to other relaxation filtering approaches, it is also shown that a larger inertial range can be obtained by the proposed anisotropic diffusion model using a compact stencil scheme in an efficient way.
\end{abstract}


\end{frontmatter}

\section{Introduction}

Most natural flow phenomena exhibit turbulence involving a considerable temporal and spatial scale separation. In order to accurately resolve all the scales of motion, it is necessary to resolve the full spectra of turbulence down to the Kolmogorov scale in a direct numerical solution (DNS). However, a fully resolved DNS requires a very fine resolution in order to avoid aliasing errors and is thus computationally prohibitive. A popular technique to reduce this computational expense is through the use of large eddy simulation (LES) in which we solve and capture the large energy containing scales and model the influence of smaller scales. In most cases, LES provides an artificial energy dissipation mechanism in order to prevent aliasing errors which manifest themselves as an energy accumulation for the higher wavenumbers. This artificial dissipation is determined by modeling the influence of the small scale structures of the flow (which are not captured by the coarse resolution) on the large scales. LES has proven to be a promising approach for the modeling of complex turbulent flows (e.g., see \cite{boris1992new,lesieur1996new,piomelli1999large,meneveau2000scale} and references therein).


The LES equations of motion for turbulent flows are derived formally by the application of a low-pass spatial filter to the governing equations. These filtered equations are considered to be a regularized form of the governing equations without the resolution requirement of small scale structures in the flow. The evolution of these equations provides a solution field for the filtered variable in space and time. However, the nonlinearity of the filtered governing equations requires the treatment of the well-known closure problem \cite{sagaut2006large,berselli2006mathematics} where the effects of the subgrid scale fluctuating eddies on the resolved scale turbulent motions are modeled. The past few decades have seen considerable efforts in the development of successful LES closure models (or simply, closures) in fluid dynamics community \cite{smagorinsky1963general,yakhot1989renormalization,yoshizawa1989subgrid,germano1991dynamic,piomelli1991subgrid,lilly1992proposed,ghosal1995dynamic,sarghini1999scale,stolz1999approximate,hughes2000large,hughes2001multiscale,winckelmans2001explicit,geurts2006leray}.

One of the earliest (and simplest) LES closures was to increase the local viscosity such that the dissipative action of the subgrid scales was approximated accurately. This added viscosity is generally called the \textit{eddy viscosity (EV)} and is the general idea behind the school of thought that prescribes functional closure models for turbulence. These models are consistent with Kolmogorov's assumption of universality for the cascade of turbulent kinetic energy from the integral length scales to the dissipation (or Kolmogorov) length scales \cite{frisch1995turbulence}. One of the most celebrated closures based on the EV hypothesis is the Smagorinsky model \cite{smagorinsky1963general} considering a functional form of turbulent eddy viscosity. This closure methodology prescribes the eddy viscosity computed from the magnitude of resolved strain rate and a characteristic length scale via a constant of proportionality (better known as the Smagorinsky constant). Although the formal derivation of the LES governing equations requires the use of spatial filter, most functional models do not explicitly specify or use one and the filter width is assumed to be related to the grid size. The Smagorinsky model has been applied to a variety of flow configurations and it is seen that the Smagorinsky constant is not fixed but rather has a wide spread of values depending on resolution, flow configuration and shear strength \cite{smagorinsky1993large,canuto1997determination,vorobev2008smagorinsky,pope2000turbulent,cushman2011introduction}. A dynamic model was proposed by Germano et al. \cite{germano1991dynamic} and improved by Lilly \cite{lilly1992proposed} where a low-pass spatial test filter was used to determine the Smagorinsky constant along with the simulation. The dynamic model has been used successfully in many fields \cite{piomelli1999large,meneveau2000scale}.

Another active area of investigation for the modeling of the dissipative action of the subgrid scale (SGS) structures is through the use of implicit LES (ILES). The basic philosophy of ILES methods is that the dissipative error of the numerical discretization schemes (generally for the nonlinear advective terms) used for the governing equations is used to account for the SGS dissipation \cite{boris1992new,hickel2006adaptive,denaro2011does}. This method is considered implicit since there is no requirement for any low-pass spatial filtering, i.e. the numerical scheme implicitly acts as a low pass spatial filter. ILES techniques are popular for their computational efficiency (as there is no turbulence model) although the control of dissipation and dispersion does tend to be difficult \cite{thornber2007implicit,grinstein2007implicit,margolin2006modeling}.


Another school of thought that addresses the concept of SGS modeling relies on the use of purely mathematical closures. These closure models are called structural closures and are based on the explicit filtering of the flow field without any prior physical assumptions or additional phenomenological arguments. Some of the more prominent structural closures include approximate deconvolution (AD), which uses the Van Cittert iterations (borrowed from the image processing community) to reconstruct the subgrid contributions to the field through the use of repeated filtering operations \cite{germano2009new,layton2012approximate,germano2015similarity}. A wide variety of flow physics have been simulated using AD-LES \cite{stolz1999approximate,domaradzki2002direct,chow2005explicit,chow2009evaluation,duan2010bridging,zhou2011large} and this method has also been mathematically analyzed in great detail \cite{dunca2006stolz,layton2006residual,layton2007similarity,rebholz2007conservation,stanculescu2008existence,dunca2011existence,berselli2012convergence,dunca2014error}. Hybrid versions of functional and structural models have been developed by coupling both concepts of eddy viscosity and low-pass spatial filtering with deconvolution \cite{habisreutinger2007coupled}. It is also common to use a separate explicit filter (in the form of an explicit or relaxation filter) as a regularization technique to account for the dissipative effect of certain subgrid scale stresses whose effects have not been captured on the resolved scales \cite{mathew2003explicit,mathew2006new}. This explicit filtering procedure can be used as a stabilization procedure for the AD-LES models or can be solely used without AD process \cite{lund2003use,bogey2004family,bogey2006large,bogey2006computation,bull2016explicit}. It is shown that the effect of multiple refiltering iterations in AD process is approximately equivalent to the combined effect of the specially designed explicit low-pass filter at the end of each time step to remove the high frequency contents near the grid cut-off scale \cite{mathew2003explicit,mathew2006new,bosshard2015udns}. Therefore, from a practical point of view, the computational cost can be reduced by the relaxation filtering framework when used without AD. On the other hand, structural models such as AD-LES and explicit filtering procedures require the careful selection of a low-pass spatial filter \cite{de2002sharp}. As such, a large number of low-pass filtering procedures have been reported in literature \cite{schumann1975subgrid,jordan1996large,najjar1996study,vasilyev1998general,sagaut1999discrete,mullen1999filtering,pruett2000priori,brandt2006priori,berland2011filter,san2015filter,san2016analysis}. It is seen, that filters with complete attenuation at the highest wavenumbers yield significantly better results in terms of the prevention of aliasing error. Both compact Pad\'{e} filters \cite{lele1992compact,visbal2002use} and selective filters \cite{bogey2004family,bogey2006large,bogey2006computation,fauconnier2013performance} have been successfully used to damp high frequency contents of the motion. In the present study, we derive another class of relaxation filters for LES using the smoothness of the flow field through the concept of anisotropic diffusion.

This document explores the use of an anisotropic diffusion closure model for solving the Burgers turbulence problem. It is generally believed that the general features of turbulence can be understood through the investigation of the Burgers equation (even though the framework of decaying Burgers turbulence is a limiting case \cite{hopf1950partial,cole1951quasi,kida1979asymptotic,frisch2001burgulence,bec2007burgers,valageas2009statistical})and it has been used to perform assessments for LES closure models due to the presence of a non-linear quadratic advective term \cite{love1980subgrid,de2002sharp,labryer2015framework,san2016analysis}. We have utilized the celebrated Perona-Malik anisotropic diffusion model \cite{perona1990scale} for the purpose of both shock-capturing (analogous to edge detection in the image processing community) and as a closure model for the SGS stresses. The Perona-Malik model and its derivatives have been used in the image processing community for image sharpening through the use of edge detection and preventing aliasing error \cite{catte1992image,yu2008noise,liang1998local,wei1999generalized}. The foundation of anisotropic diffusion lies in the unequal dissipation of certain areas in the solution field as a function of the local spatial gradients in that area. This ability to diffuse selectively is used in preserving high gradient areas (such as the edges or shocks) and dissipating any possible oscillations in the smoother areas. We aim to use this ability to prevent oscillations in smoother areas as a method to damp the amplitude of dispersion errors from the underlying numerical schemes. An excellent discussion of the general idea of anisotropic diffusion may be found in \cite{weickert1998anisotropic}.

The proposed closure model can be considered analogous to a relaxation filtering framework except that the dissipation of the dispersion errors is carried out through the addition of an artificial viscosity term through one iteration of a parabolic equation. Through this, it is proposed that a better preservation can be obtained for the inertial range which a great majority of relaxation filters sacrifice to prevent aliasing errors. Nonlinear anisotropic diffusion has previously been implemented for the purpose of filtering and edge enhancement for numerical methods related to conservation laws (specifically the nonlinear higher order dissipative terms) such as the Lax-Wendroff formula \cite{grahs2002image}, oscillatory central schemes \cite{grahs2002entropy} and the shock capturing effect on the inviscid Burgers equation \cite{breuss2005stabilised,wei2002shock}. In addition to testing the general framework of anisotropic diffusion as a technique for LES closures, different diffusivity kernels (i.e. heuristics that alter the application of the anisotropic diffusion) are also tested to assess their robustness. We also note that the proposed anisotropic diffusion filtering scheme is constructed using a compact stencil locally consisting of only three points. An effective damping mechanism for the high frequency flow structures can be obtained by using this 3-point stencil scheme via defining nonlinear diffusivity kernels. The interested reader is directed to Keeling et al. \cite{keeling2002nonlinear} for an in-depth discussion on the characteristics of these diffusivity kernels. The relaxation filtering procedure based on the 9-point stencil selective filter is used for comparison purposes.

The layout of this paper is as follows. The governing equations and the background numerical methods are reported in Section 2. Section 3 describes the proposed Perona-Malik anisotropic diffusion method and a set of diffusivity kernels studied in this investigation. The shock-capturing feature and SGS modeling ability of the method are documented through a sensitivity analysis for a single mode sine wave and the Burgers turbulence case (using 32 sample fields) in Section 4. Concluding remarks are outlined in Section 5.

\section{Governing equation and numerical methods}

\newcommand{\partfrac}[2]{\frac{\partial #1 }{\partial #2}}
\newcommand{\partfracsec}[2]{\frac{\partial^2 #1 }{\partial #2^2}}

The viscous Burgers equation is considered a test bed representing a one-dimensional (1D) homogeneous flow and is a good starting point for the evaluation of a numerical method before proceeding to the three-dimensional (3D) Navier-Stokes equations. The evolution of the velocity field $u(x,t)$ in the Burgers equation is given by

\begin{align}
    \frac{\partial u}{\partial t} + u \frac{\partial u}{\partial x} = \nu \frac{\partial ^2 u}{\partial x^2},
\end{align}
where $\nu$ is the kinematic viscosity. The quadratic nonlinearity of the advective term mimics a transport mechanism and the Laplacian term behaves as a dissipative mechanism. For this reason, the 1D viscous Burgers equation has been popular for novel SGS model development \cite{kida1979asymptotic,love1980subgrid,gotoh1993statistics,bouchaud1995scaling,blaisdell1996effect,gurbatov1997decay,balkovsky1997intermittency,de2002sharp,adams2002subgrid,bec2007burgers,labryer2015framework}.

In this work, we have chosen to use the conservative form of the Burgers equation. A general representation of this form can be given by
\begin{align}
    \partfrac{u}{t} + R(u) = L(u),
    \label{BurgEq}
\end{align}
where $R(u)$ and $L(u)$ are the nonlinear and linear operators given by
\begin{align}
\begin{gathered}
    R(u) = \frac{1}{2} \partfrac{u^2}{x},\\
    L(u) = \nu \partfracsec{u}{x}.
\end{gathered}
\end{align}
Other formulations for the nonlinear term are described in detail in Blaisdell et al. \cite{blaisdell1996effect}. The spatial derivatives in the formulation ($R(u)$ and $L(u)$) were discretized by using a sixth-order central compact difference schemes \cite{lele1992compact} to ensure minimal truncation error. The compact scheme for the first order derivative is given by
\begin{align}
    \frac{1}{3}f'_{j-1} + f'_{j} +  \frac{1}{3} f'_{j+1} = \frac{14}{9}\frac{f_{j+1}-f_{j-1}}{2h} + \frac{1}{9}\frac{f_{j+2}-f_{j-2}}{4h}\, ,
    \label{eq:com1}
\end{align}
where the superscript prime signifies the first derivative and $h$ is the uniform spatial discretization length. Eq.~(\ref{eq:com1}) can be solved by the Thomas algorithm \citep{press1992numerical} to give us an approximation of the first derivative. The second order derivatives were computed in a similar manner by
\begin{align}
    \frac{2}{11} f''_{j-1} + f''_{j} +  \frac{2}{11} f''_{j+1} = \frac{12}{11}\frac{f_{j+1}-2f_{j} + f_{j-1}}{h^2} + \frac{3}{11}\frac{f_{j+2}-2f_{j} +f_{j-2}}{4h^2} \, ,
\label{eq:com2}
\end{align}
where $f^{''}$ denotes the second derivative.

A system of semi-discrete ordinary differential equations (ODEs) are obtained after spatially discretizing our governing partial differential equation (PDE) using the compact schemes explained above following which our system can be represented by
\begin{align}
    \partfrac{u_j}{t} = \pounds (u_j)
\end{align}
where the spatial gradient terms are denoted in the $\pounds(u_j)$ term as follows
\begin{align}
    \pounds(u_j) = -R(u_j) + L(u_j).
\end{align}

The third-order accurate in time, total variation diminishing Runge-Kutta scheme (TVDRK3) \cite{gottlieb1998total} is used to integrate our discrete system of equations in time. The integration scheme is described in the following, where it has been assumed that the time level $l$ is known from which we aim to estimate our solution at time level $l+1$.

\begin{align}
    \label{eq:TVDRK}
    \begin{split}
    u^{(1)}_j & = u^{\ell}_j + \Delta t \pounds(u^{\ell}_j)  \\
    u^{(2)}_j  & = \frac{3}{4}  u^{\ell}_j + \frac{1}{4} u^{(1)}_j + \frac{1}{4}\Delta t \pounds (u^{(1)}_j)  \\
    u^{\ell+1}_j  & = \frac{1}{3}  u^{\ell}_j + \frac{2}{3} u^{(2)}_j + \frac{2}{3}\Delta t \pounds (u^{(2)}_j).
    \end{split}
\end{align}
This work uses a time step that ensures no errors related to time integration of the system of ODEs obtained using the above described method of lines. Both single mode sine wave and the Burgers turbulence problems are simulated over a domain of $x\  \epsilon \  [0,2 \pi]$, with periodic boundary conditions using $\nu = 5 \times 10^{-4}$.

\section{Perona-Malik anisotropic diffusion}

Widely used in image processing, Perona-Malik diffusion (also known as anisotropic diffusion) is a methodology used to reduce image noise without compromising the quality of certain significant parts of the image content such as edges or lines which may be important for the interpretation of the image. It is similar to the blurring techniques which employ the use of an isotropic Gaussian filter with parameters controlling the filter width. However, isotropic Gaussian filters are considered linear and space-invariant in their transformation behavior and do not depend on the local content of the original (noisy) image. Anisotropic diffusion produces a parameterized image which is a combination of the original image and a filter that prescribes a blurring operation depending on the local content of the original image making the process space variant but \textit{nonlinear}. In the famous Perona-Malik model \cite{perona1990scale}, the filter kernel is isotropic (in that it applies the same filter at different areas of the image field) but allows for lower dissipation at the edges or other significant structures. Anisotropic diffusion is implemented through the use of a generalized diffusion equation, i.e., each new (or filtered) image is obtained by applying this diffusion equation to the previous image. It must be noted here that, although the technique is named \textit{anisotropic} diffusion, the filter kernel is actually isotropic since it utilizes a scalar-valued diffusivity instead of a direction-dependent diffusion tensor \cite{weickert1997review}. Another interesting feature of this technique is the union of filtering and edge detection processes in a single procedure contrary to most algorithms in the image processing community which utilize two independent processes to be applied in series for the same outcome. As an analogy to fluid dynamics, the filtering procedure to remove noise from images can be considered akin to removing dispersion errors propagating from errors due to the finite difference approximations of discontinuities. And edge detection algorithm is similarly applicable for the capture of discontinuities in the solution field such as those formed by shocks.

The Perona-Malik technique is performed after each complete time integration procedure. In this method, the obtained field values at discrete grid points are passed through a parabolic equation which adds dissipation according to certain rules of diffusivity. This can be clarified by detailing the partial differential equation that adds diffusion
\begin{align}
    \frac{\partial u}{\partial \tau} = \nabla c. \nabla u + c(x) \Delta u,
\end{align}
where $\tau$ is the pseudo-time. Increased iterations of the above equation lead to higher dissipation added to the field value. The number of iterations required for adding an adequate amount of diffusion is problem specific and a considerable amount of research is still underway in that field. An interesting point to make here is that the use of a constant kernel for $c$ would be akin to specifying a Gaussian blurring operation over the entire solution field.

In this study, the number of pseudo-time iterations is kept at a default value of one iteration to mimic an explicit filtering mechanism. This is intended to ensure that the induced dissipation does not exceed the requisite amount. This also ensures that the total dissipation is controlled by the pseudo-time step and the diffusion constant. The above continuous equation can be discretized in the following manner
\begin{align}
    \label{PMeqa}
    u^*_i = u_i + \frac{\Delta \tau}{2 h^2} \left\{(c_i + c_{i+1})(u_{i+1}-u_{i})- (c_{i-1}+c_{i})(u_i-u_{i-1})\right\}
\end{align}
or re-scaling $\tau$ it can be written as
\begin{align}
    \label{PMeq}
    u^*_i = u_i + \frac{\kappa \Delta t}{2 h^2} \left\{(c_i + c_{i+1})(u_{i+1}-u_{i})- (c_{i-1}+c_{i})(u_i-u_{i-1})\right\}
\end{align}
where $c_i$ is our diffusivity function varying as a function of location, $h$ is the spatial grid discretization and $\kappa$ is the ratio between pseudo-time step ($\Delta \tau$) and physical simulation time step ($\Delta t$).  We note here that the value of $\kappa$ must be very small to ensure that an inordinate amount of dissipation is not added to the solution field (by way of a large pseudo-time step).

There are several diffusivity functions in use for the purpose of edge detection (in image processing applications) and three popular kernels used for edge detection have been tested in this work. We have, as proposed originally by Perona-Malik
\begin{align}
    \label{CondFunc}
    c(x) = \frac{1}{1+\left|\frac{\nabla u}{K}\right|^2},
\end{align}
where $K$ is the diffusion scaling parameter. In the image processing methodology, $K$ represents the sensitivity of the isotropic filter kernel to edges and lines in the solution field. In our study, the gradient term, $\nabla u=\frac{\partial u}{\partial x}$, will be available at the end of the time step.
A variation of the above kernel is provided by \cite{geman1992constrained}
\begin{align}
    \label{CondFunc1}
    c(x) = \frac{1}{\left(1+\frac{1}{3}\left(\frac{\nabla u}{K}\right)^2\right)^2}.
\end{align}
Structurally it can be seen that the above kernel behaves similarly to the kernel originally proposed by Perona and Malik and we expect the general trend of the dissipation characteristics for both kernels to be similar.

A third conductivity, also proposed in the initial work of Perona and Malik is tested which uses an exponential type kernel to impart dissipation to the solution field as against the reciprocal approach described above. The kernel is given by
\begin{align}
    \label{CondFunc2}
    c(x) = \exp{\left(-\frac{1}{2}\left(\frac{\nabla u}{K}\right)^2\right)}.
\end{align}
The above presentation of kernels has been done in the increasing order of decay rates. For instance, the exponential kernel is seen to reduce the magnitude of applied dissipation rapidly with increasing absolute values of gradients. It can be inferred that the exponential function given by Eq.~(\ref{CondFunc2}) is more localized in its dissipation characteristics with higher gradient values unaffected by the kernel as compared to the other kernels.

\begin{figure}[!t]
\centering
\mbox{
\subfigure[Kernel 1 given by Eq.~(\ref{CondFunc})]{\label{KernelTFsa}\includegraphics[width=0.5\textwidth,trim=4 4 6 6,clip]{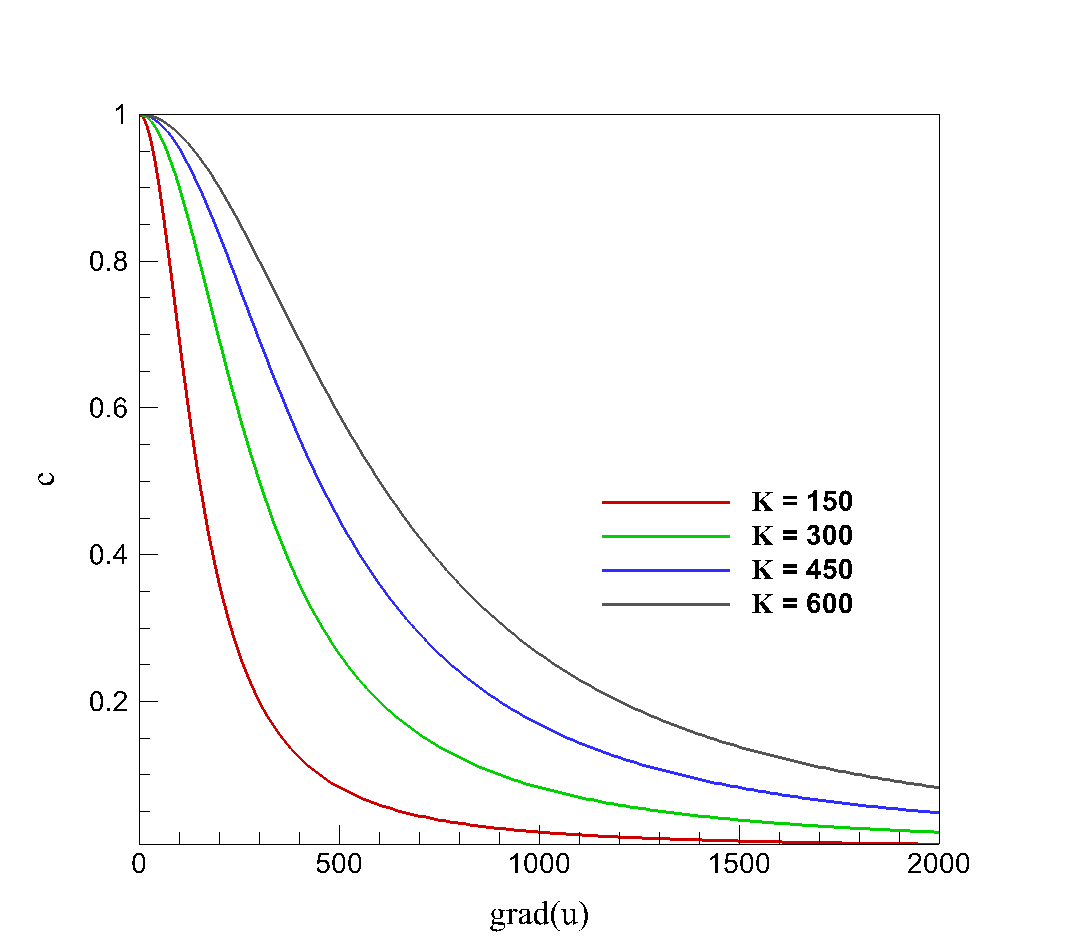}}
\subfigure[Kernel 2 given by Eq.~(\ref{CondFunc1})]{\label{KernelTFsb}\includegraphics[width=0.5\textwidth,trim=4 4 6 6,clip]{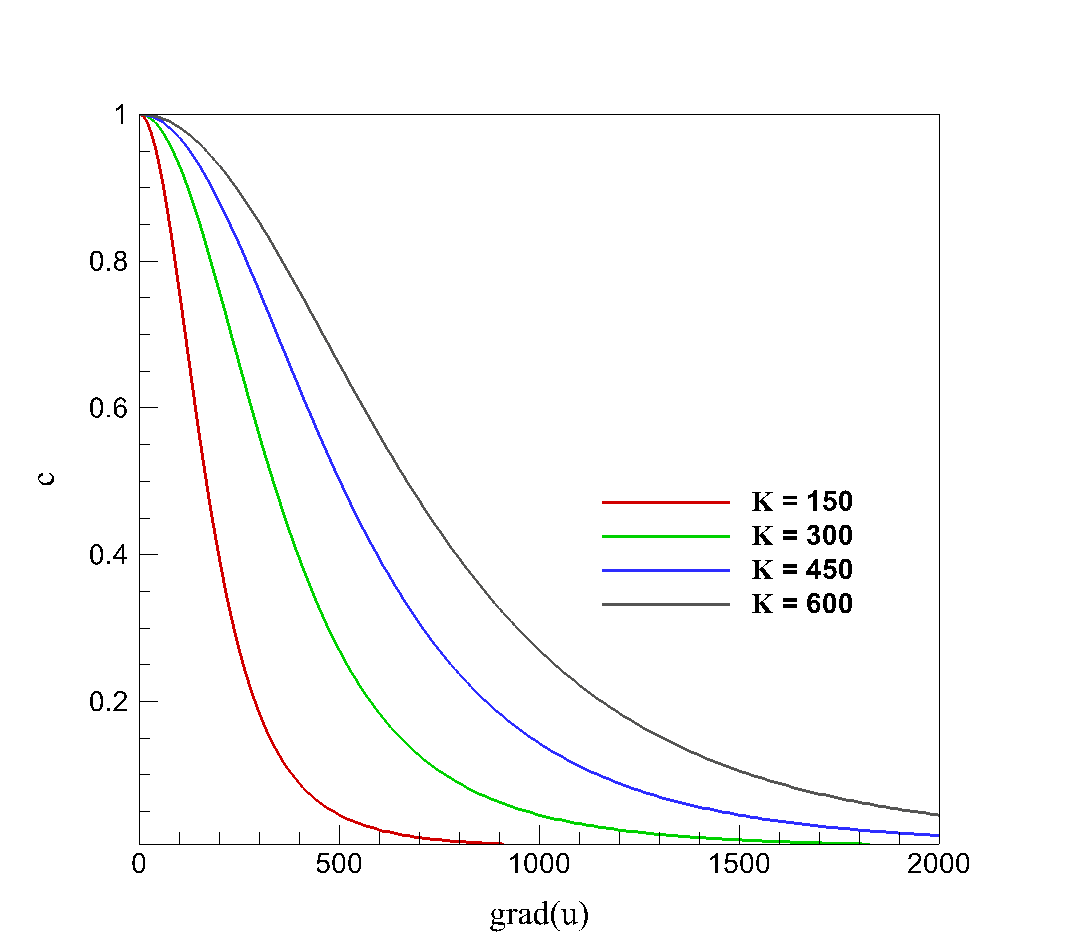}}}
\mbox{
\subfigure[Kernel 3 given by Eq.~(\ref{CondFunc2})]{\label{KernelTFsc}\includegraphics[width=0.5\textwidth,trim=4 4 6 6,clip]{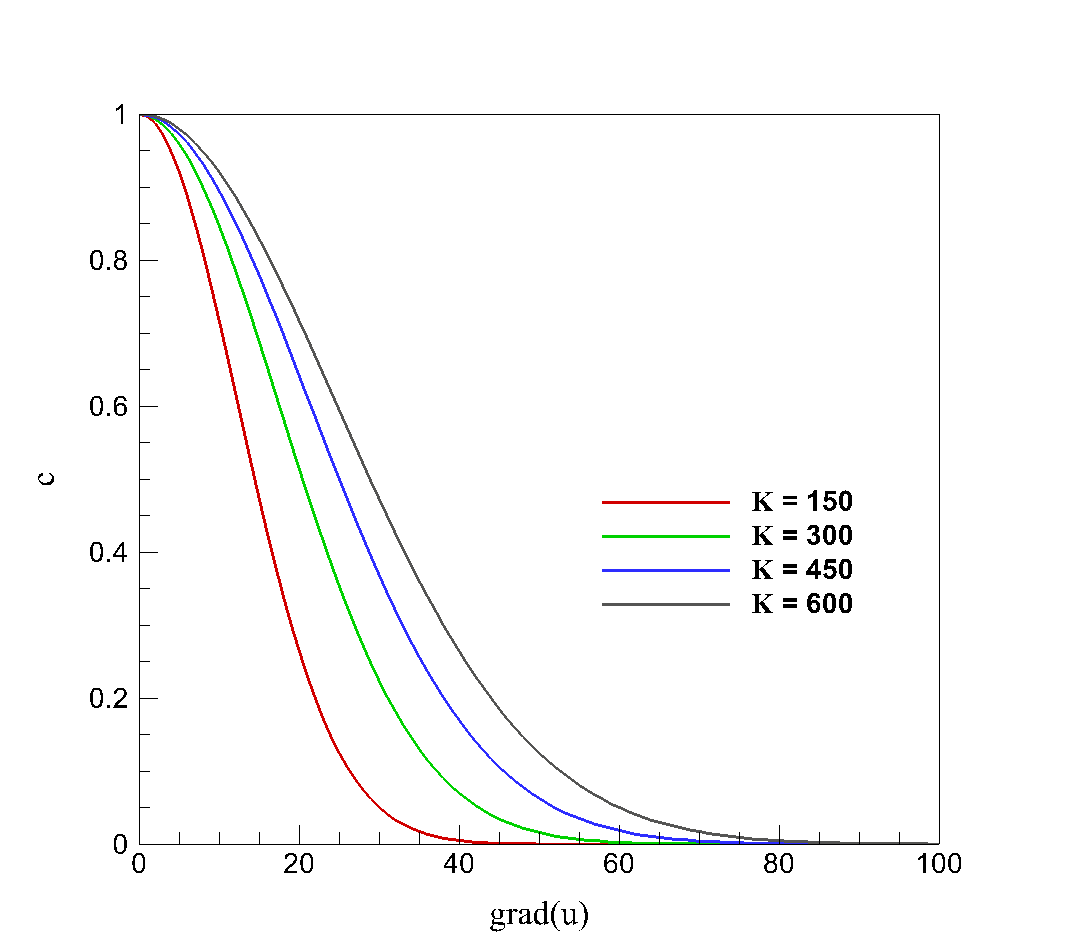}}
}
\caption{Dissipative behavior of diffusivity kernels}
\label{KernelTFs}
\end{figure}

One must observe here that the structures of these kernels were designed to cause a similar amount of dissipation in the limit of very large or very small values of the diffusion constant $K$, but their behavior in the intermediate ranges varied slightly. The Fig.~\ref{KernelTFs} shows the behavior of the kernel for varying one dimensional gradient values (indicative for the Burgers turbulence case). Counter to intuition, the kernels apply an anisotropic type of diffusion by dissipating locations with \textit{lower} absolute values of the gradient while preserving high gradient regions. This characteristic of the kernels is key in the preservation of edges and the damping out of any possible grid to grid oscillation propagation in the smoother regions. We mention here that the proposed framework allows for the potential use of customized kernels for diffusivity and slightly different dissipation characteristics but the broader trends elaborated in this research still hold true. The extension of the aforementioned concepts to higher dimensions is straightforward.

The proposed framework is tested against the well known relaxation filtering framework proposed by Bogey et al \cite{bogey2004family}. The 9-point stencil explicit filtering scheme is given by
\begin{align}\label{bogey}
    u^*_i = u_i - \sigma_d D_i\ \  \textnormal{where}\ \ D_i = \sum_{j = -4}^{4} d_j u_{i+j}.
\end{align}
The coefficients $d_j$ are given by
\begin{align}
        d_0 = 35/128, \quad d_1 = -7/32, \quad d_2 = 7/64, \quad  d_3 = -1/32, \quad d_4 = 1/256,
\end{align}
and the parameter $\sigma_d$ ($0 \leq \sigma_d \leq 1$) is used for dissipation control with higher values of $\sigma_d$ providing lower dissipation.


\section{Results}

This section covers the results of the shock capturing performance of the proposed closure model as well as the ability to capture the theoretical Burgers turbulence scaling for the different test cases used. Using a standard LES methodology, coarse resolution LES runs are compared to fully resolved DNS data (obtained from higher resolutions). Conclusions are then drawn about the efficacy of the proposed closure. A sensitivity analysis is also carried out to determine the effect of the modeling parameters in the closure. All results for the proposed model are compared to under-resolved DNS (UDNS) data at the same (i.e., coarse) resolution to assess the gain through LES.

\begin{figure}[!t]
\centering
\mbox{
\subfigure[UDNS]{\label{Figure1a}\includegraphics[width=0.5\textwidth,trim=4 4 6 6,clip]{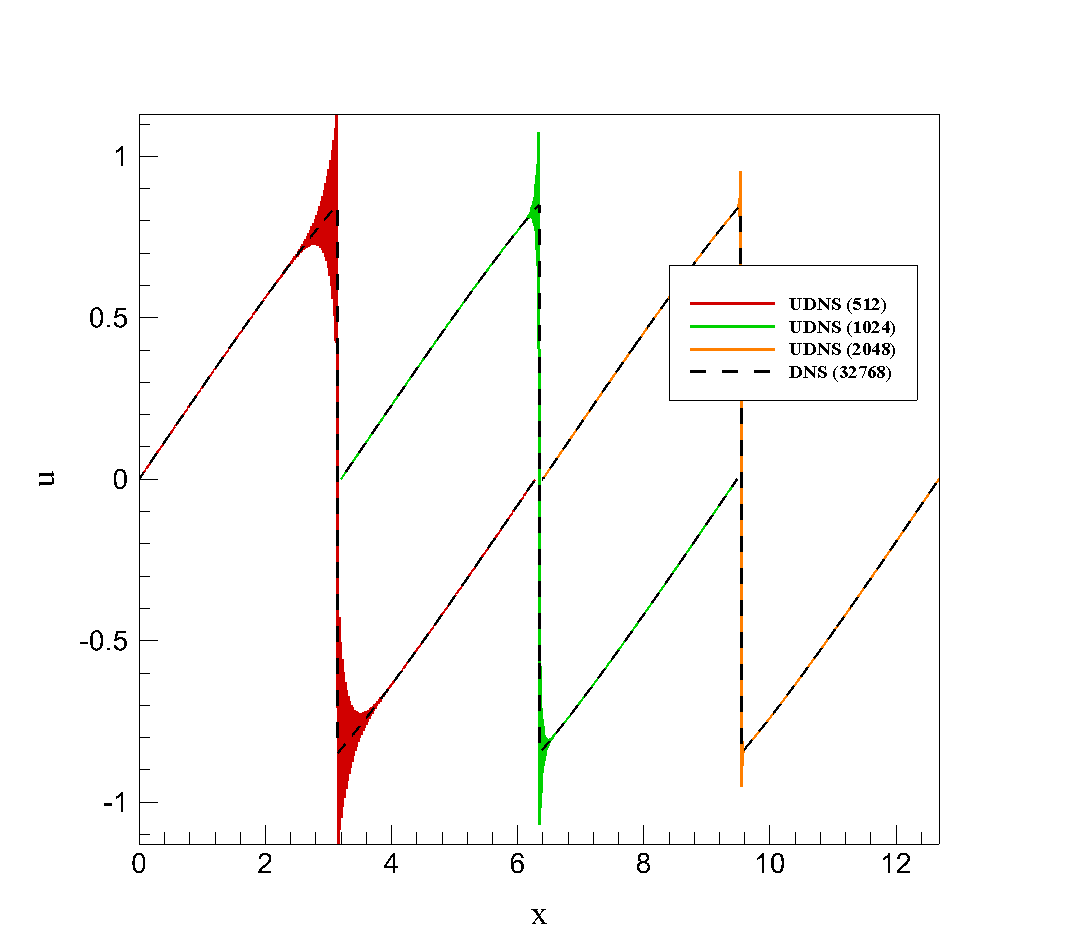}}
\subfigure[Anisotropic diffusion]{\label{Figure1b}\includegraphics[width=0.5\textwidth,trim=4 4 6 6,clip]{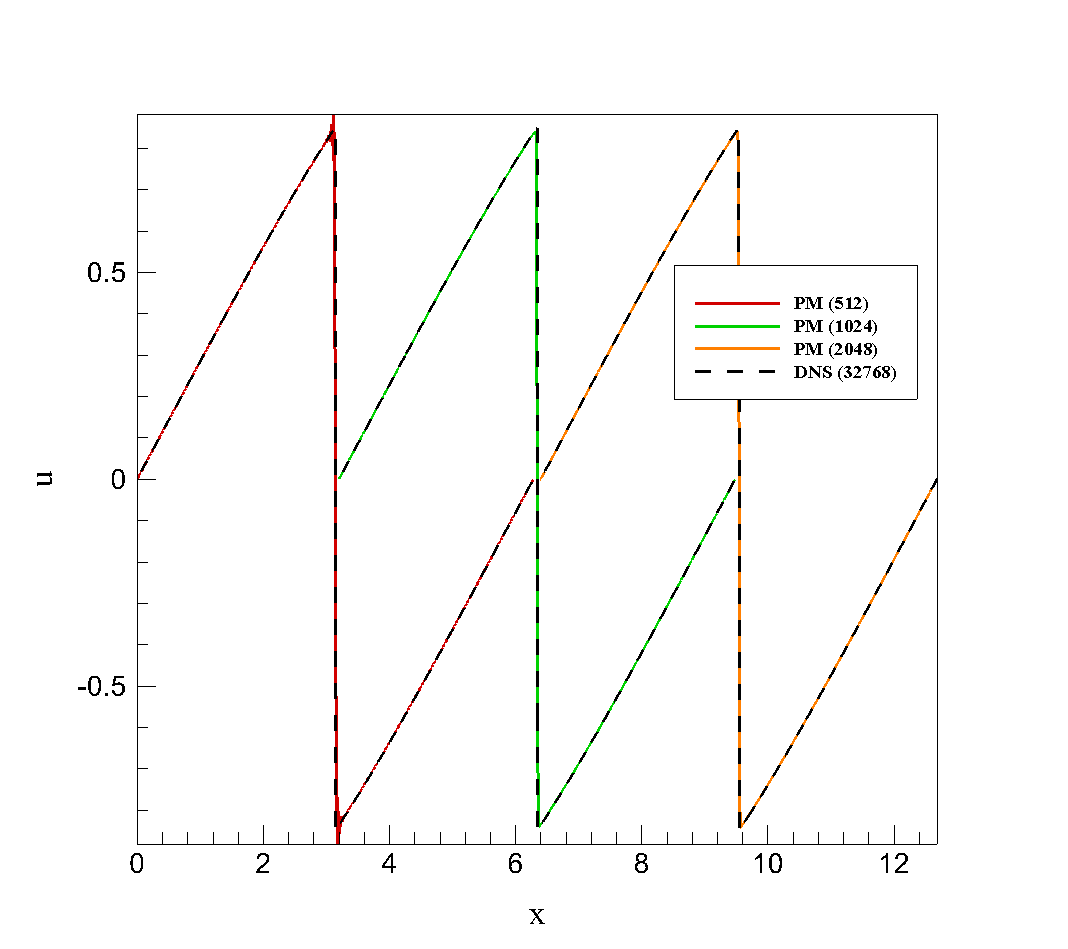}}
}
\caption{Shock capturing ability of the Perona-Malik anisotropic diffusion model with $K = 200$ and $\kappa = 0.0015$ at resolutions of $N=512$, $1024$ and $2048$.}
\label{Figure1}
\end{figure}

First, we present the behavior of a shock formulation problem initiated by a single-mode single Sine wave $u(x,0) = \sin{x}$ as an initial condition. Fig.~\ref{Figure1} shows the simulation results (at time $t = 2.5$) for this case demonstrating shock formation at a time approximately equal to 1.5. For the purpose of comparison $N=32768$ resolution DNS runs are also presented. Coarse grid simulations of $N=512$, $N=1024$ and $N=2048$ are shown for both UDNS and LES runs and it is evident that the UDNS runs (as illustrated in Fig.~\ref{Figure1}\subref{Figure1a}) show a large number of grid-to-grid oscillations particularly for the coarsest resolution. However it is seen that the proposed model demonstrates a remarkable ability to damp out the higher Fourier modes propagating from the discontinuities due to the dispersion error inherent to central schemes as can be seen in Fig.~\ref{Figure1}\subref{Figure1b}.

As mentioned previously, the decaying Burgers turbulence problem is considered in a domain of $x\ \epsilon [0,2 \pi]\ $ with periodic boundary conditions. The initial conditions of the problem examined to determine closure performance are prescribed by an initial energy spectrum:
\begin{align}
    E(k) = A k^4 \exp{-(k/k_0)^2},
\end{align}
and $A$ is a constant given by
\begin{align}
    A = \frac{2 k_0^{-5}}{3 \sqrt{\pi}}.
\end{align}
The above conditions ensure a total energy $\int E(k) dk = 1/2$ at the start of the simulation. The parameter $k_0$ is assumed to be 10 and corresponds to the wavenumber where one obtains the maximum scale of the initial energy spectrum. Thus, we can generate velocity magnitudes in Fourier space from this initial energy spectrum as
\begin{align}
    \left| \hat{u} (k) \right|= \sqrt{2 E(k)}.
\end{align}
Ensemble averaged simulations are carried out using different realizations of the initial energy spectrum generated by
\begin{align}
    \left| \hat{u} (k) \right|= \sqrt{2 E(k)}\exp{i 2 \pi \Psi(k)},
\end{align}
where $\Psi(k)$ is a uniform random number distribution between 0 and 1 at each wavenumber. The aforementioned, random number distribution also satisfies a conjugate relationship given by $\Psi(k) = -\Psi(-k)$ to obtain a real velocity field in physical space. A Fast Fourier transform algorithm \cite{press1992numerical} is used for the purpose of inversion to and from Fourier space. 32 randomly selected sample fields are constructed with the aid of the random number distribution and simulated for different phases. The simulations are terminated once the energy content of the flow becomes significantly lower than the initial value. In the following, the results from these ensemble averaged runs are computed and presented. Identical random number seeds are used to ensure that the initial conditions for all studies presented are same. The quantity of interest to be investigated is the energy spectrum given by
\begin{align}
    E(k,t) = \frac{1}{2} \left|\hat{u}(k,t)\right|^2,
\end{align}
which gives a total energy of
\begin{align}
    E(t) = \int_{-k_m}^{k_m} E(k,t) dk.
\end{align}

\begin{figure}[!t]
\centering
\mbox{
\subfigure[Energy spectra]{\label{Figure2a}\includegraphics[width=0.5\textwidth,trim=4 4 6 8,clip]{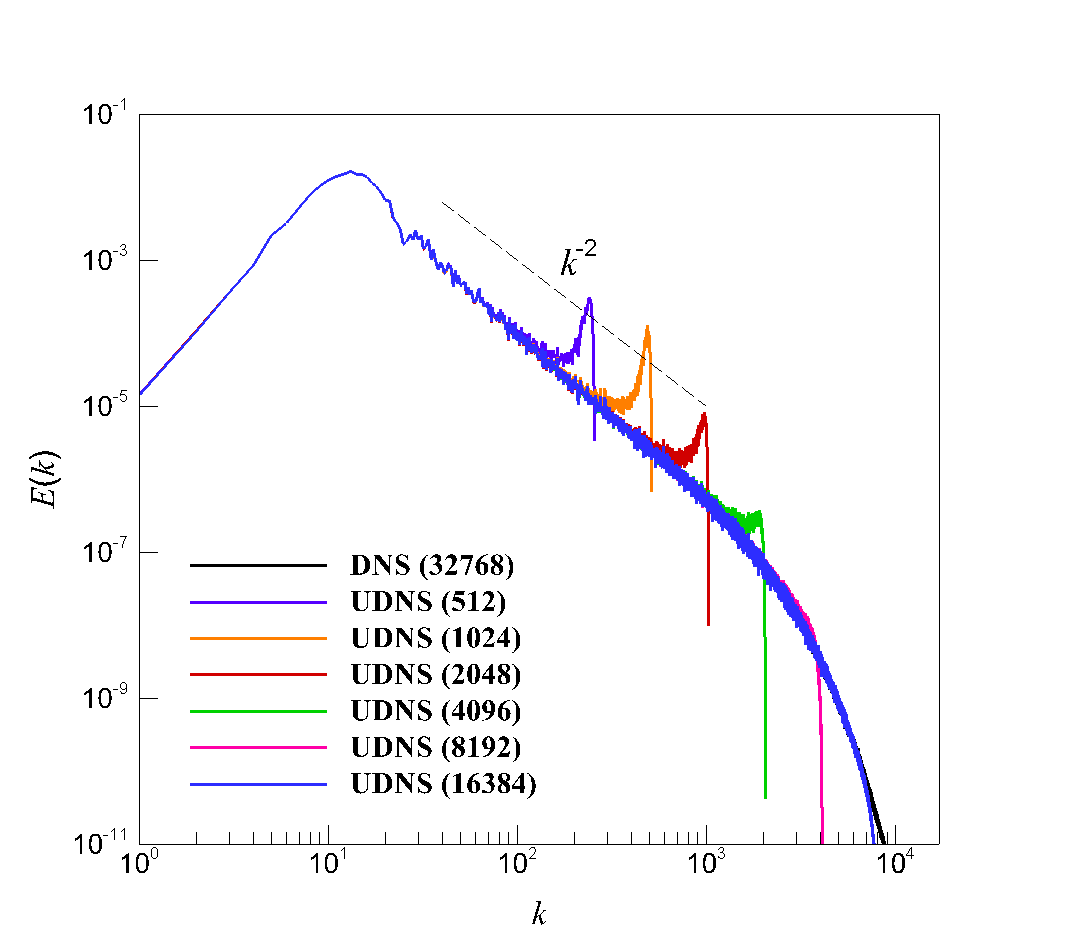}}
\subfigure[Dissipation rates]{\label{Figure2b}\includegraphics[width=0.5\textwidth,trim=4 4 6 8,clip]{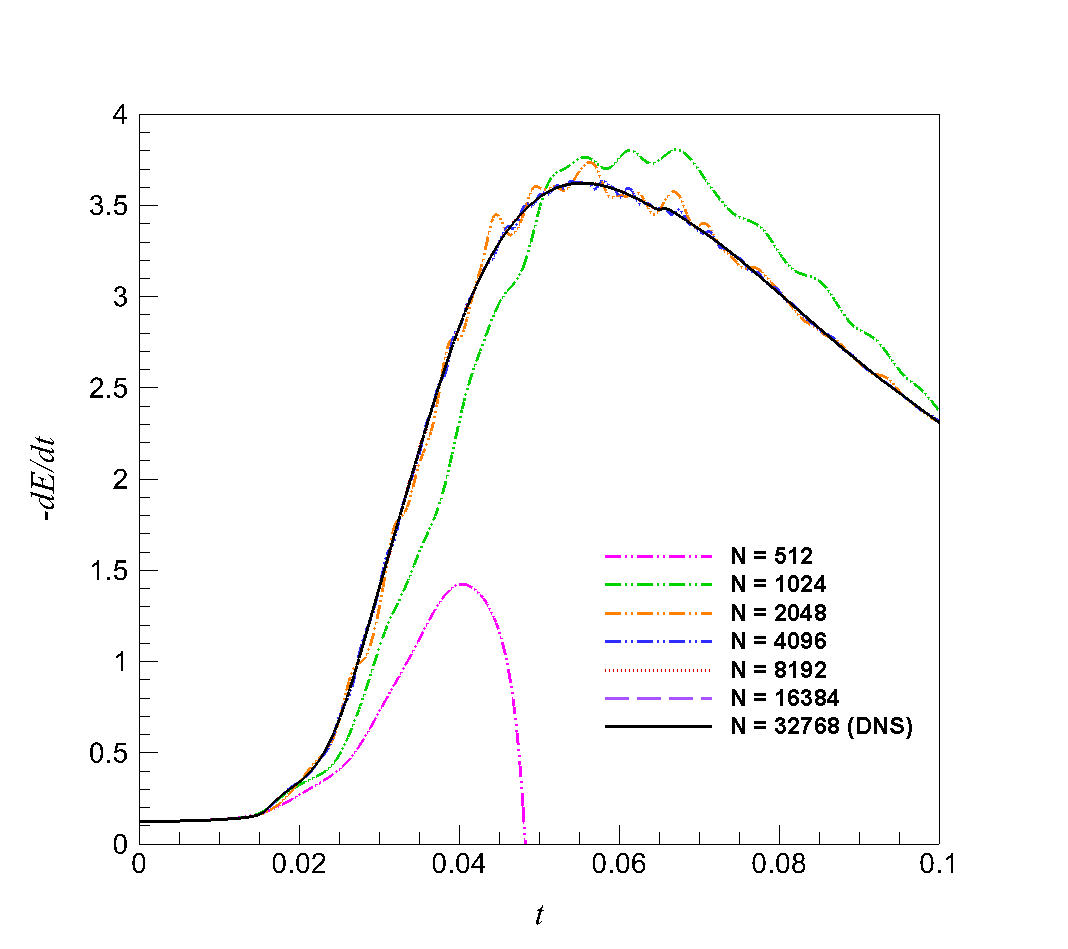}}
}
\caption{Comparison of energy spectra (at time $t = 0.05$) and time evolution of dissipation rates for the decaying Burgers turbulence problem at $\nu = 5 \times 10^{-4}$ with varying grid resolution from very coarse mesh ($N = 512$) to fully resolving DNS mesh ($N = 32768$). Ideal scaling $k^{-2}$
is also included in the energy spectra plot.}
\label{Figure2}
\end{figure}

The performance of numerical methods can also be analyzed using other measures and the reader is directed to \cite{san2016analysis} for further information. A grid convergence study is carried out to determine the DNS resolution requirements for the chosen viscosity (i.e., $\nu = 5 \times 10^{-4}$). Time integration is carried out till $t=0.1$s and we note here that the shocked solution is obtained at $t=0.05$s which also corresponds to the highest energy dissipation rate as can be seen in Fig.~\ref{Figure2b}. Resolutions of $N=512$ to $N=32768$ are tested and it is observed that the highest resolution corresponds to a high-fidelity DNS capturing all the scales in the problem without aliasing error. The grid convergence study can be seen in Fig.~\ref{Figure2a}. Higher resolutions are seen to eliminate grid-to-grid oscillations gradually in the form of a reduced accumulation of energy at the wavenumbers corresponding to the grid cut-off. The DNS also shows the theoretical scaling for the Burgers turbulence energy spectrum with increasing wavenumbers (i.e., $k^{-2}$). The coarsest resolutions (512, 1024 and 2048) are chosen for our closure modeling assessments due to a significant presence of energy accumulation at those resolutions.

The dissipation control kernels proposed in this work have two free parameters: $K$ and $\kappa$. Through a sensitivity analysis it is aimed that the contribution of each modeling parameter be quantified in terms of aliasing error prevention and preservation of inertial range scaling. Different resolutions are used to ensure consistency. For the purpose of comparison, UDNS and DNS results are also plotted alongside the sensitivity analysis. At the end of this section, an indicative figure of a well-known relaxation filtering scheme (the 9-point stencil selective filter) is included for the purpose of comparison. It is seen that a considerable improvement in the range of inertial scales is obtained. Default parameters for the sensitivity analysis are kept at $K = 125$ and $\kappa = 0.0015$ and their values are systematically varied.

\begin{figure}[!t]
\centering
\mbox{
\subfigure[$N=512$]{\label{Figure3a}\includegraphics[width=0.5\textwidth,trim=4 4 6 6,clip]{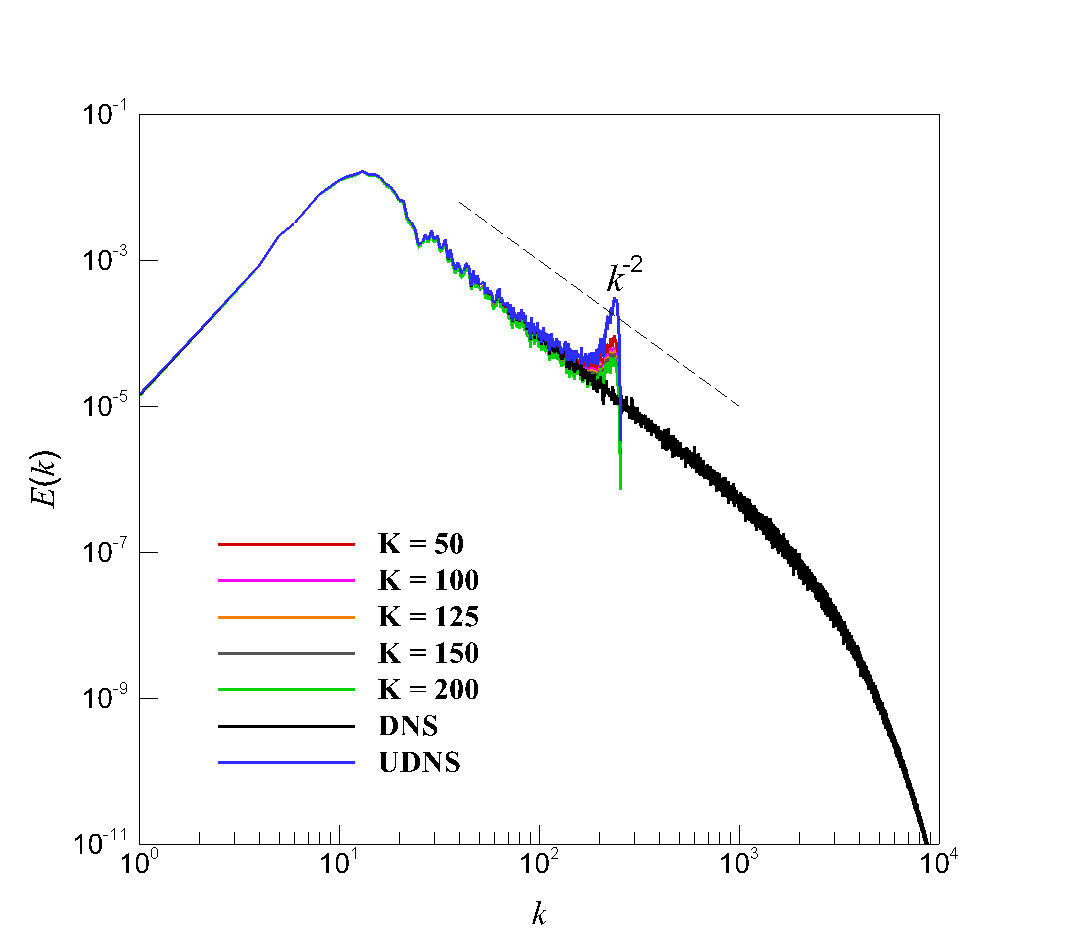}}
\subfigure[$N=1024$]{\label{Figure3b}\includegraphics[width=0.5\textwidth,trim=4 4 6 6,clip]{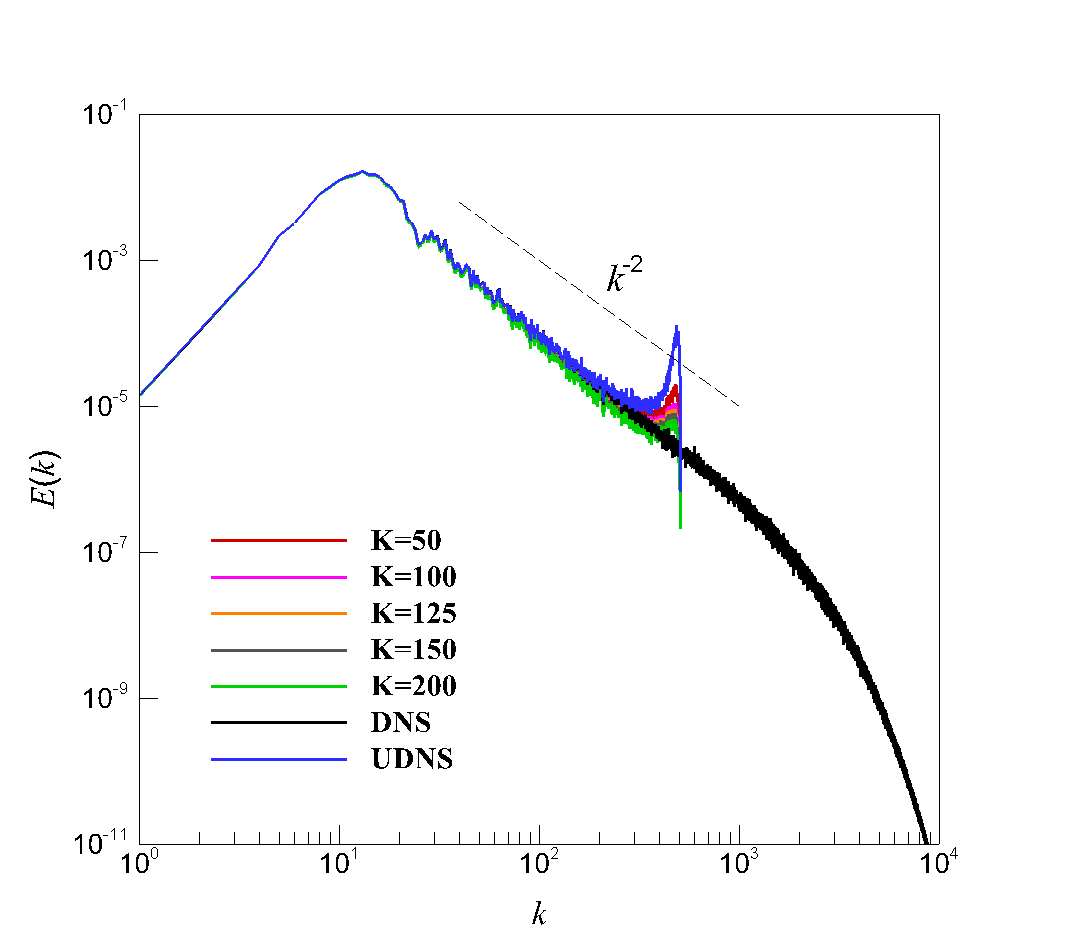}}}
\mbox{
\subfigure[$N=2048$]{\label{Figure3c}\includegraphics[width=0.5\textwidth,trim=4 4 6 6,clip]{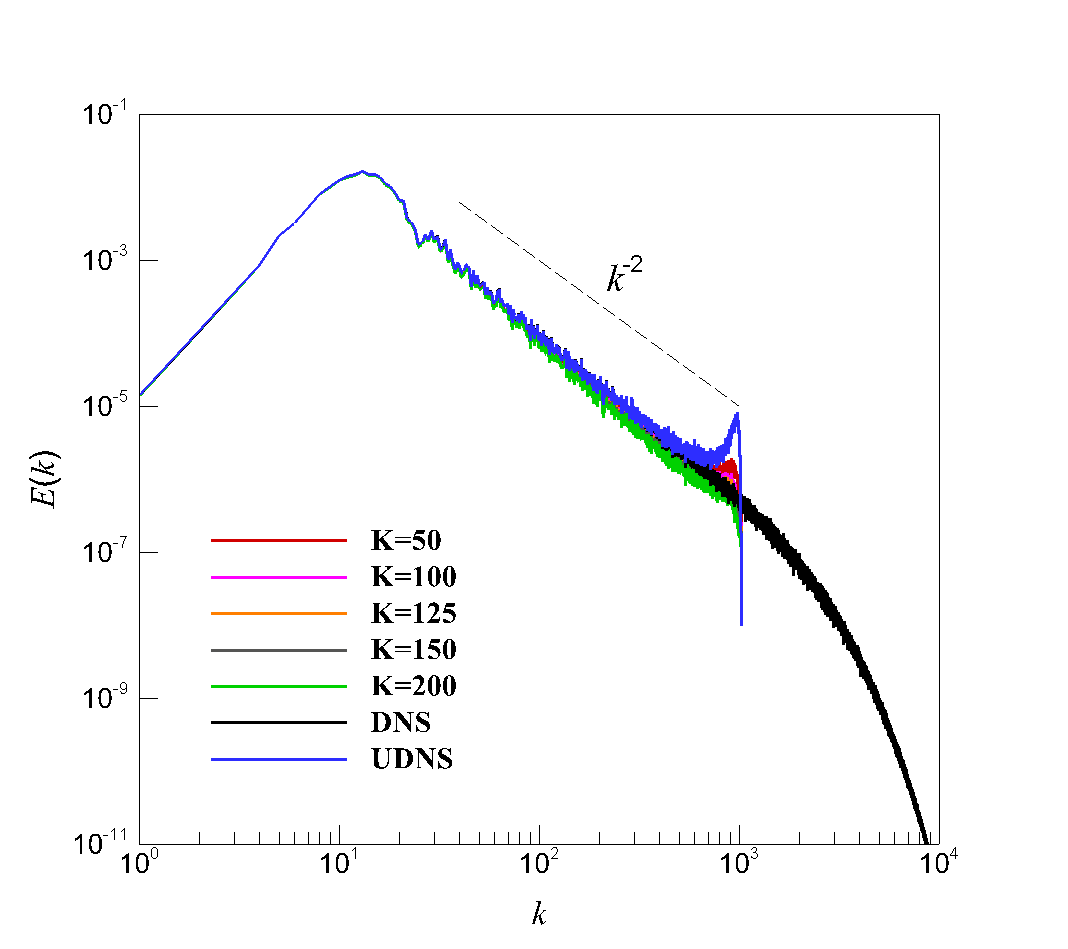}}
}
\caption{Effect of diffusion constant $K$ for diffusivity kernel 1.}
\label{Figure3}
\end{figure}

\begin{figure}[!t]
\centering
\mbox{
\subfigure[$N=512$]{\includegraphics[width=0.5\textwidth,trim=4 4 6 6,clip]{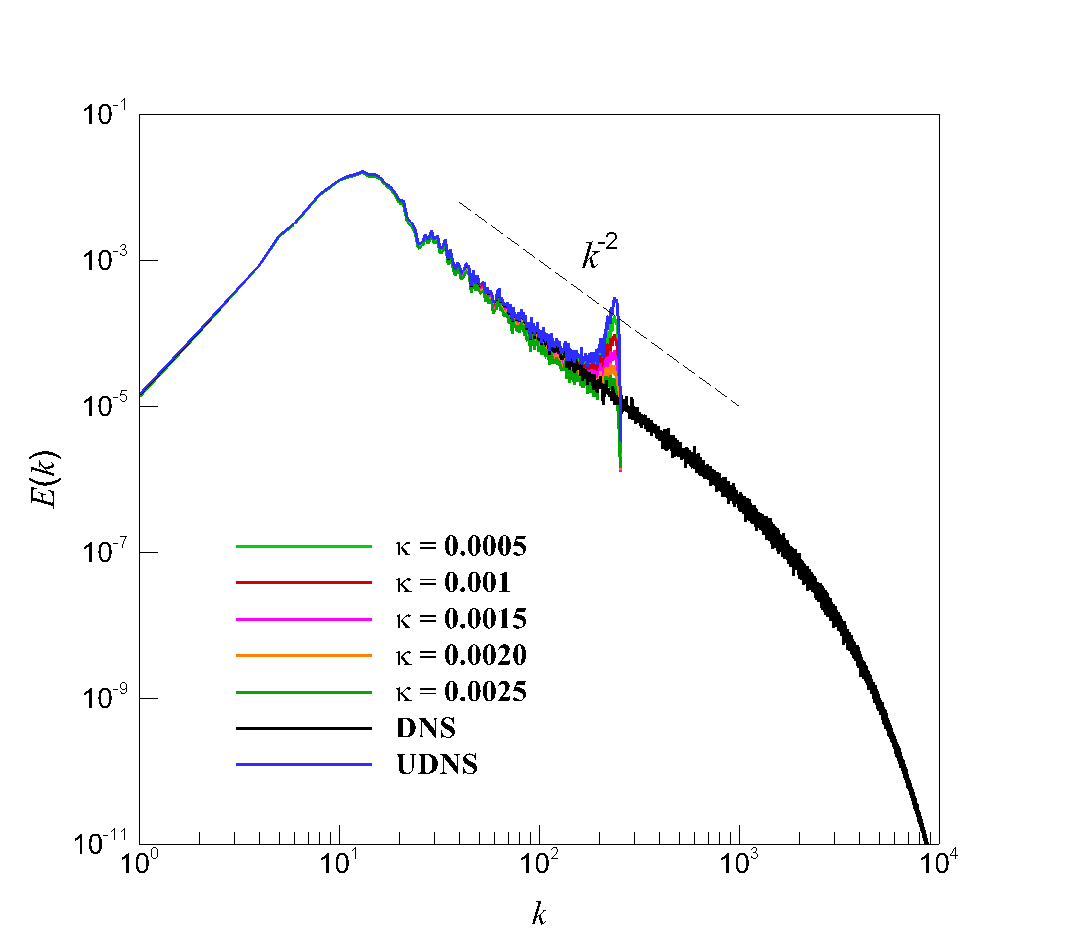}}
\subfigure[$N=1024$]{\includegraphics[width=0.5\textwidth,trim=4 4 6 6,clip]{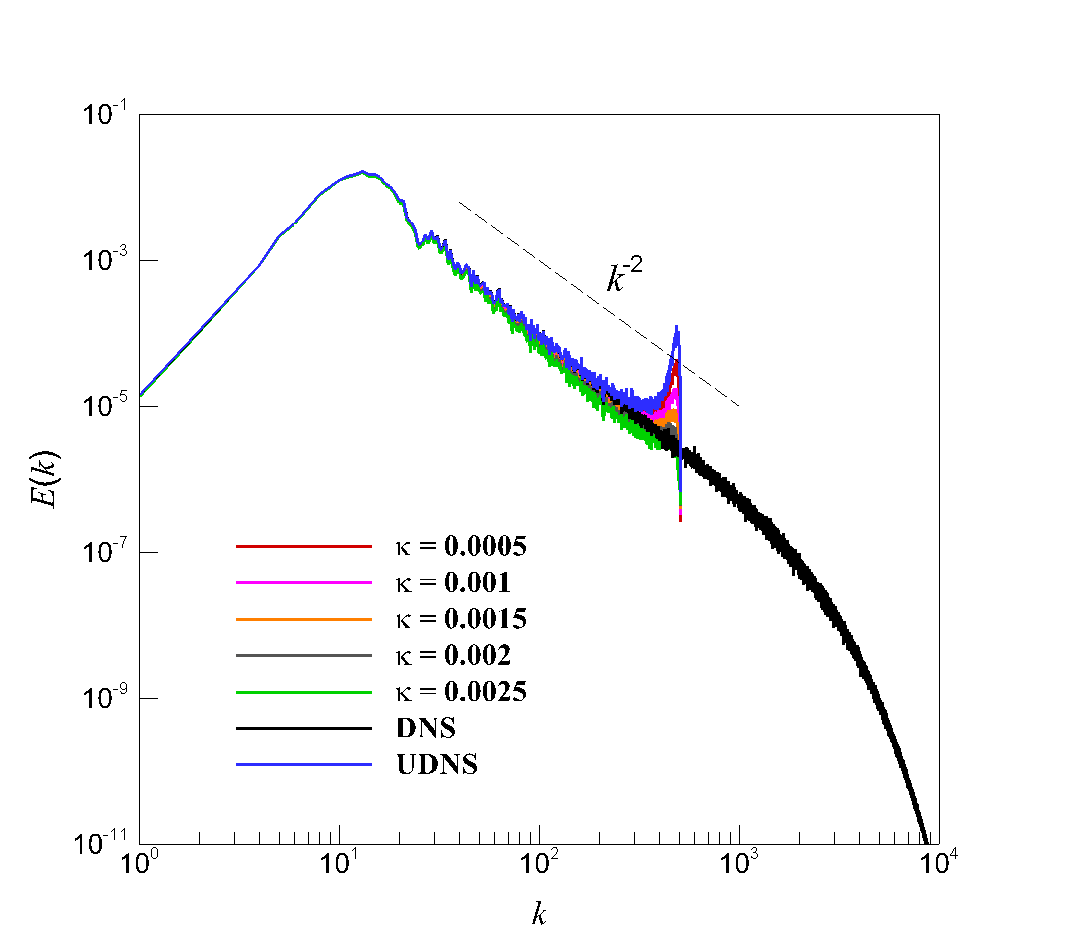}}}
\mbox{
\subfigure[$N=2048$]{\includegraphics[width=0.5\textwidth,trim=4 4 6 6,clip]{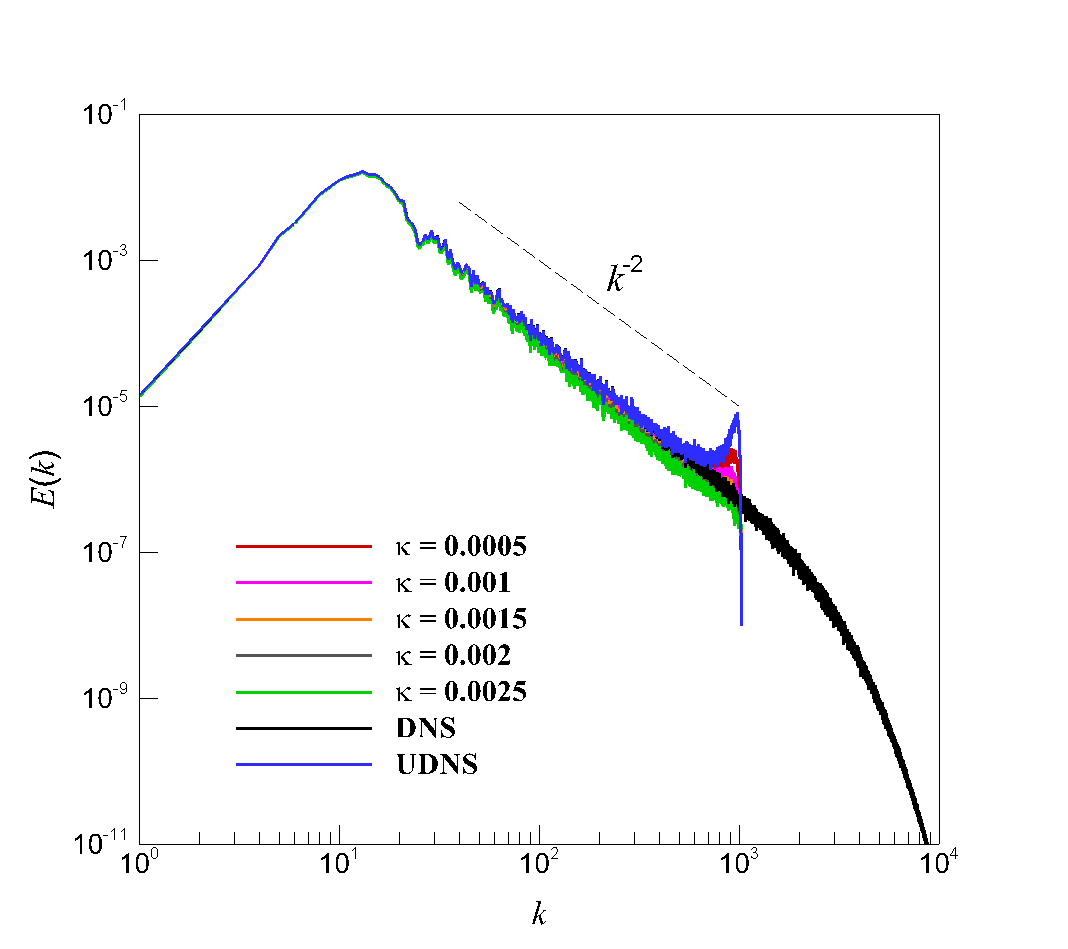}}
}
\caption{Effect of $\kappa$ for diffusivity kernel 1.}
\label{Figure4}
\end{figure}

Fig.~\ref{Figure3} shows the effect of the diffusion scaling constant $K$ on the conductivity kernel 1 (given by the kernel in Eq.~(\ref{CondFunc})). The increase in this constant causes an increased reduction of the energy accumulation at grid cut-off. It is noted here that, with increase in $K$ the effect of the dissipation gradually increases to a limit beyond which the kernel values are equal for all gradients (i.e., $c = 1$ for all $\nabla u$). Fig.~\ref{Figure4} shows the effect of $\kappa$ (the ratio of physical time step to pseudo-time step) on the energy spectrum for the coarse simulations. Increasing values of $\kappa$ cause increased dissipation due to a higher pseudo-time step. The aforementioned figures indicate that a slight adjustment of parameters is required to ensure an adequate amount of dissipation. While extremely accurate scaling can be obtained even for the coarsest of resolutions ($N=512$), the same set of parameters may prove more dissipative for the higher resolutions.

\begin{figure}[!t]
\centering
\mbox{
\subfigure[$N=512$]{\label{Figure5a}\includegraphics[width=0.5\textwidth,trim=4 4 6 6,clip]{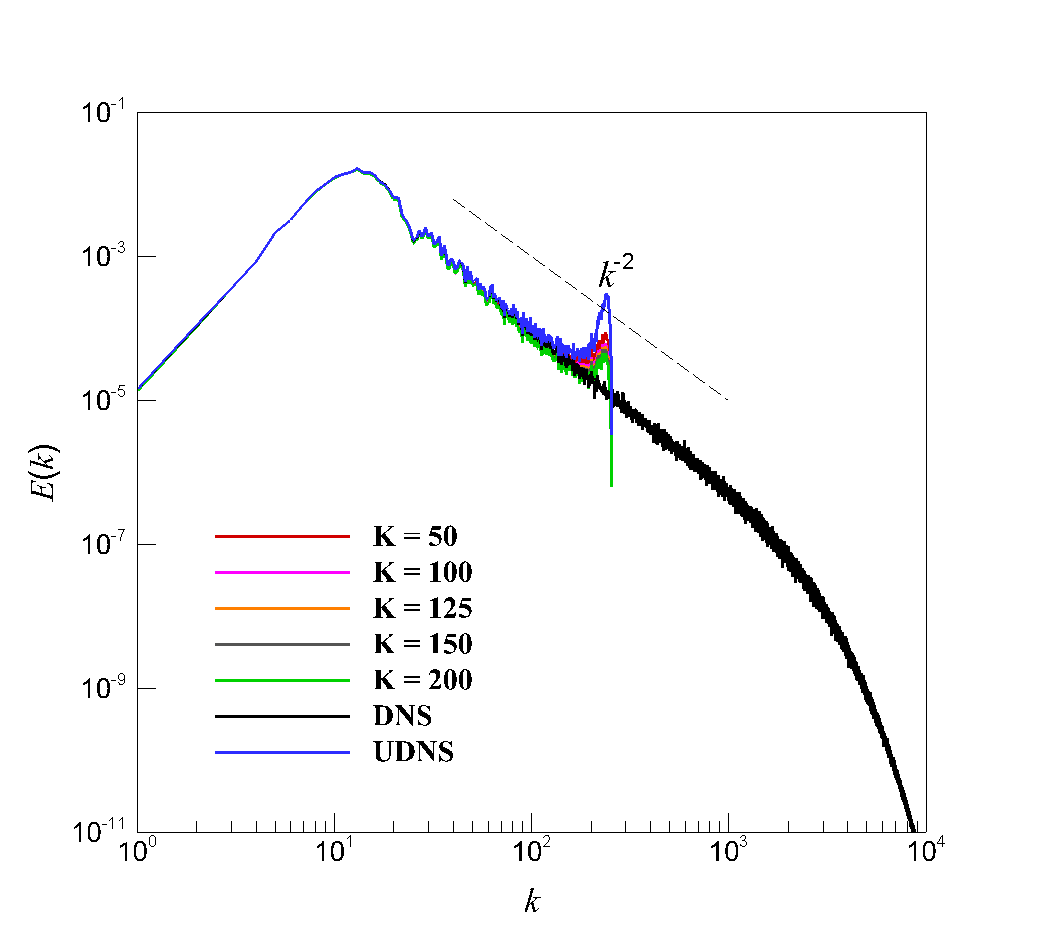}}
\subfigure[$N=1024$]{\includegraphics[width=0.5\textwidth,trim=4 4 6 6,clip]{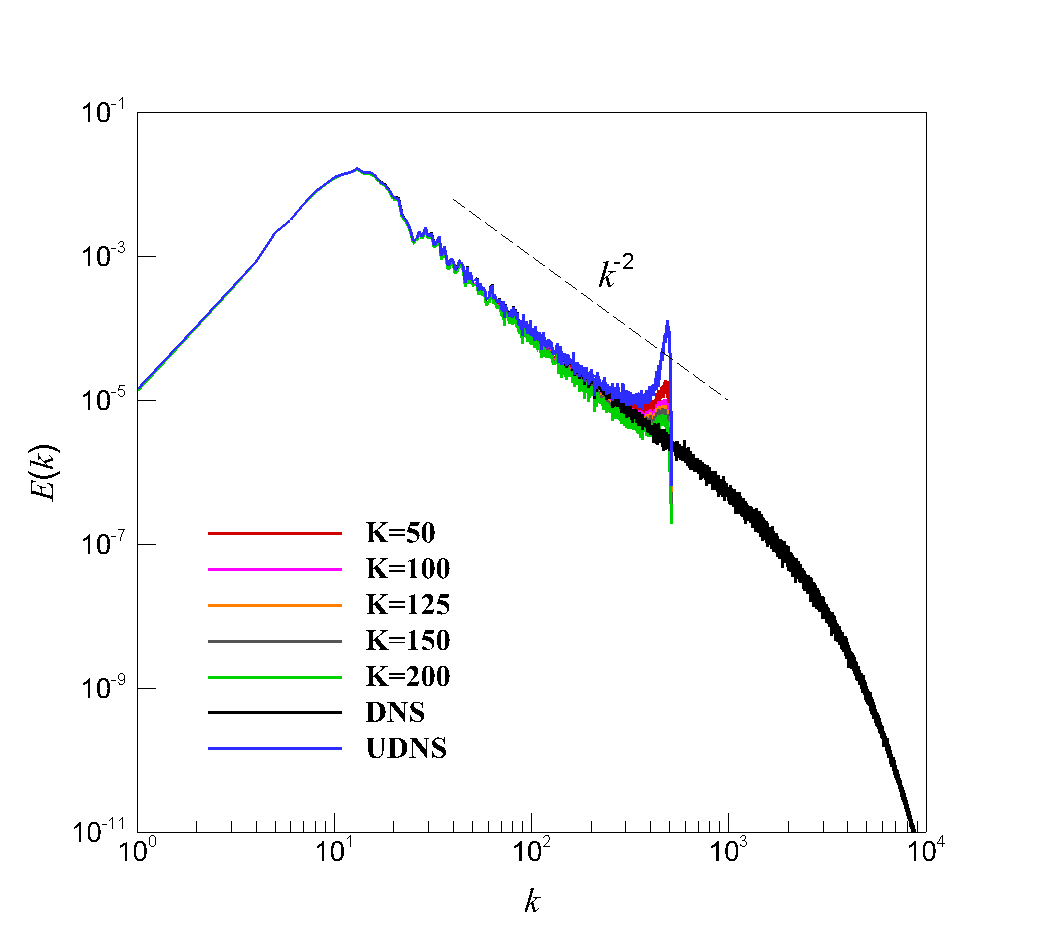}}}
\mbox{
\subfigure[$N=2048$]{\label{Figure5b}\includegraphics[width=0.5\textwidth,trim=4 4 6 6,clip]{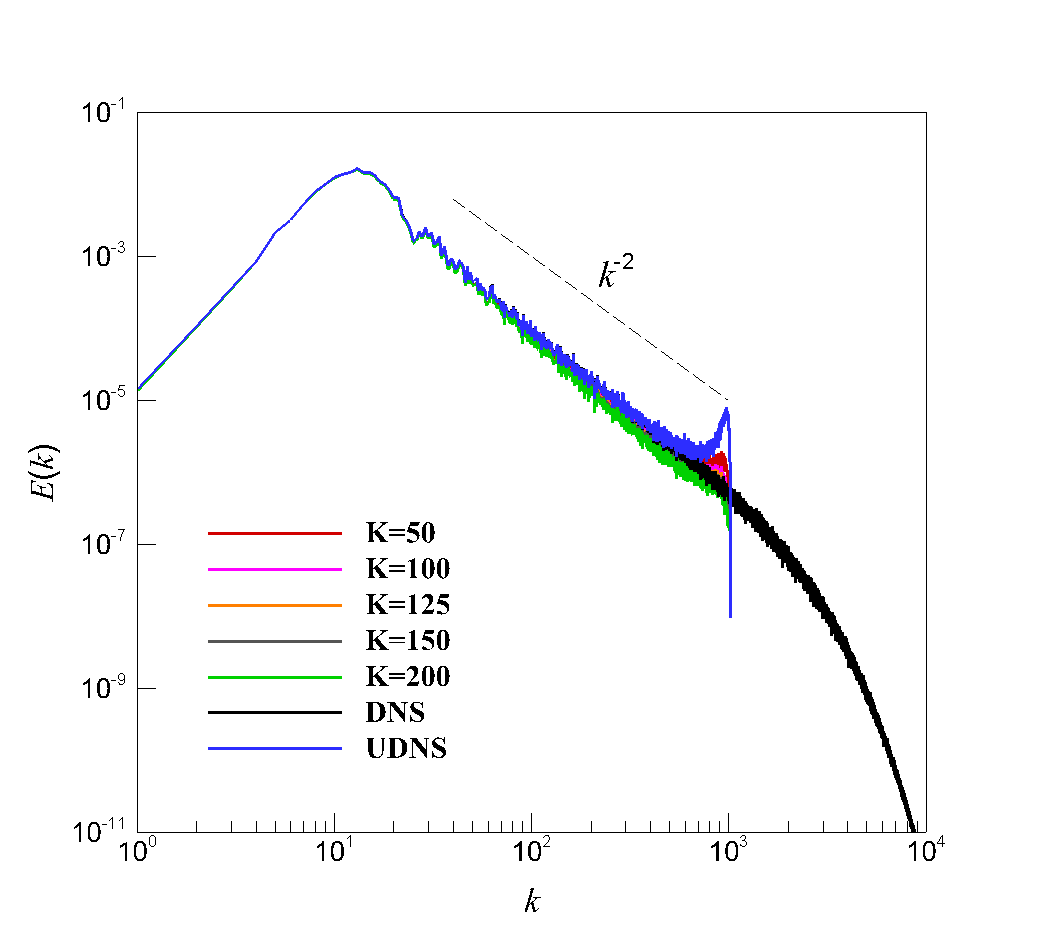}}
}
\caption{Effect of diffusion constant $K$ for diffusivity kernel 2.}
\label{Figure5}
\end{figure}

\begin{figure}[!t]
\centering
\mbox{
\subfigure[$N=512$]{\includegraphics[width=0.5\textwidth,trim=4 4 6 6,clip]{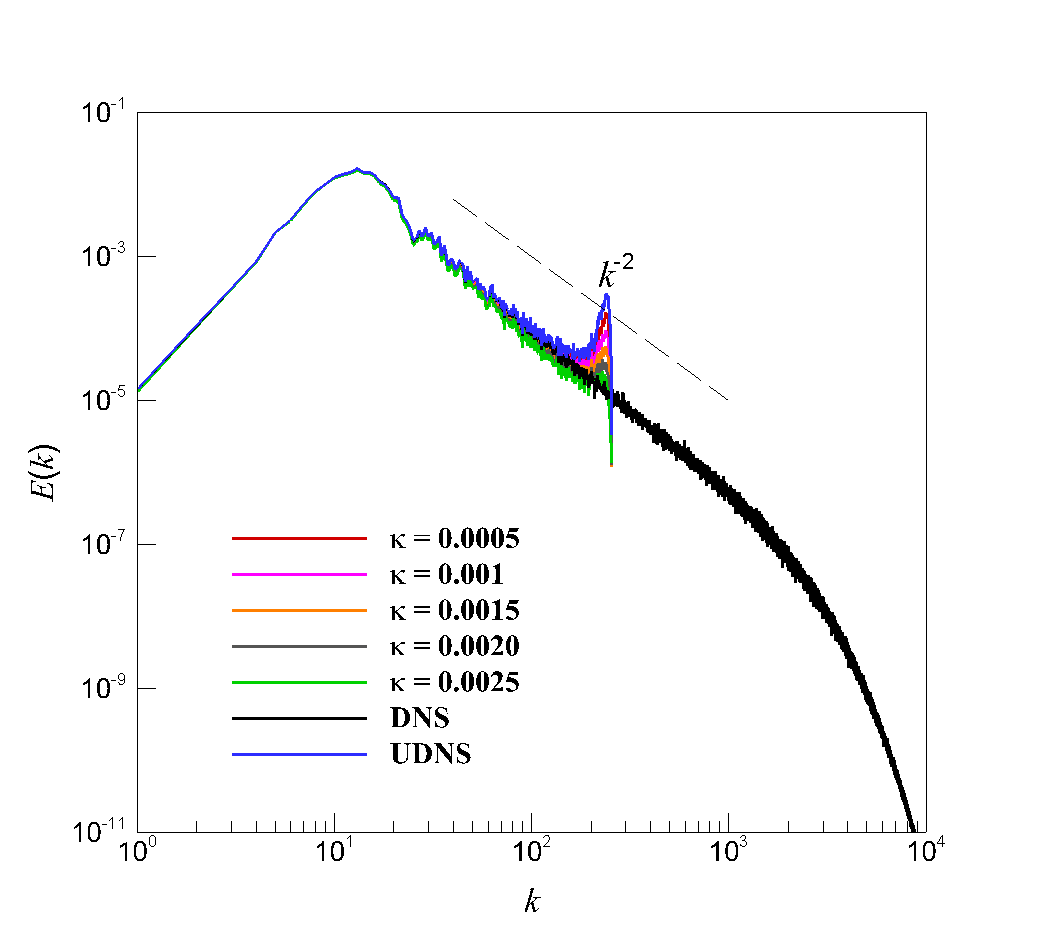}}
\subfigure[$N=1024$]{\includegraphics[width=0.5\textwidth,trim=4 4 6 6,clip]{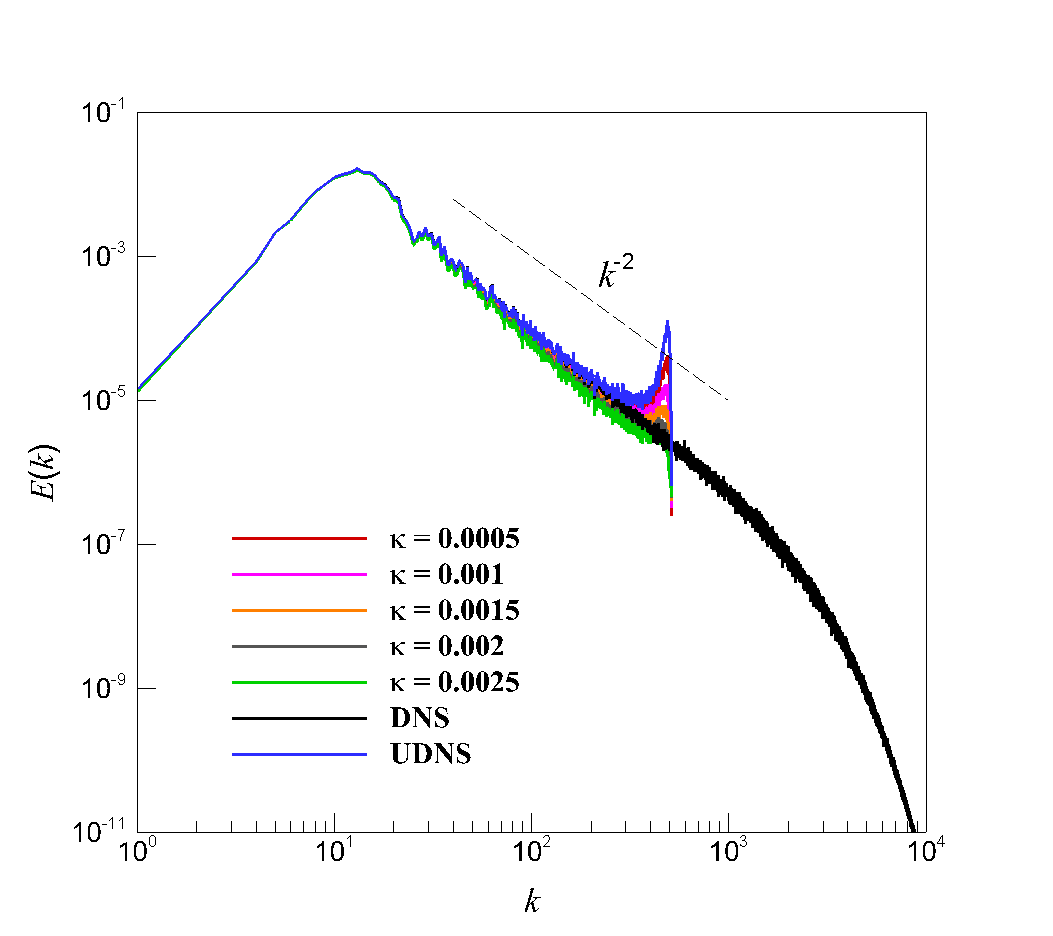}}}
\mbox{
\subfigure[$N=2048$]{\includegraphics[width=0.5\textwidth,trim=4 4 6 6,clip]{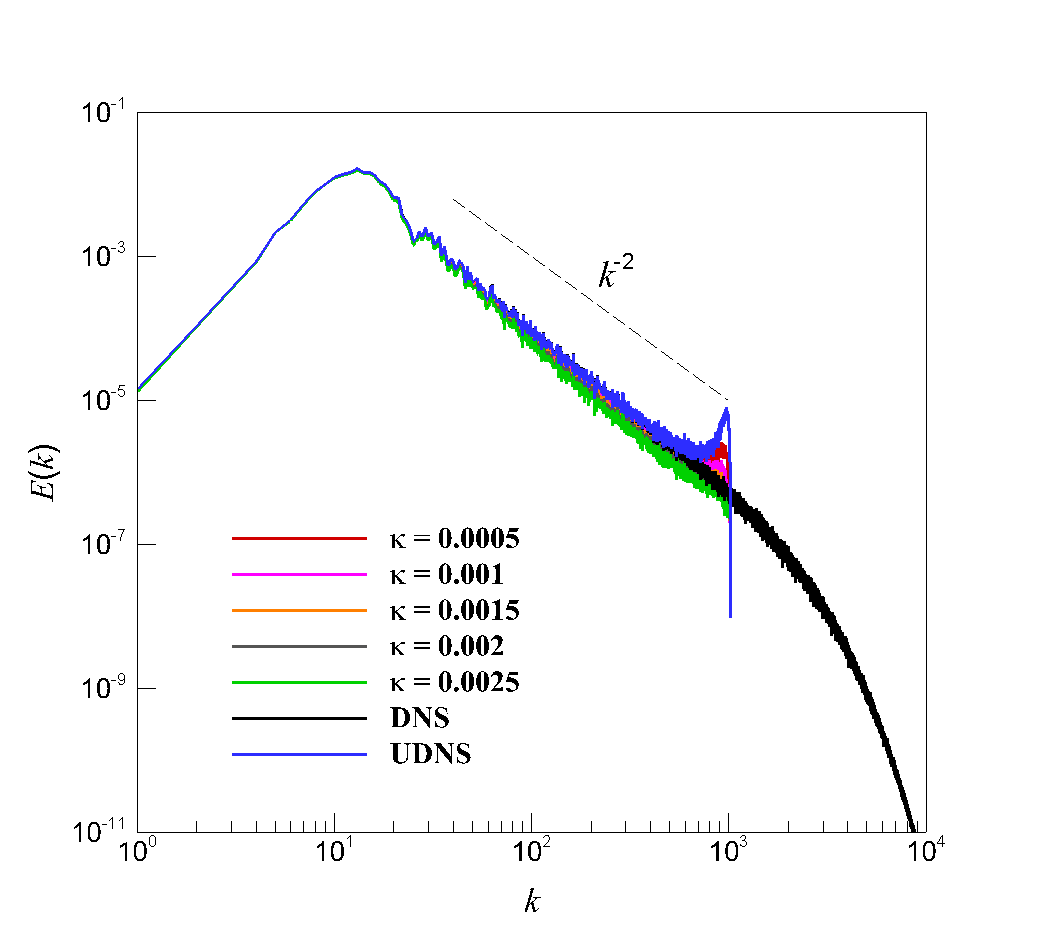}}
}
\caption{Effect of $\kappa$ for diffusivity kernel 2.}
\label{Figure6}
\end{figure}

Fig.~\ref{Figure5} and  Fig.~\ref{Figure6} show the effect of our dissipation control parameters for the conductivity kernel specified by Eq.~(\ref{CondFunc1}). As expected, the behavior of these scaling parameters was fundamentally similar to the first kernel due to their analogous structures. This can be verified by the fact that their kernels show the same dissipation behavior with increasing gradient values (Fig.~\ref{KernelTFsb}) and that the limiting dissipation ($c = 1$) is approached at almost identical rates. It must be noted here that we have fixed the number of iterations of the parabolic equation given by Eq.~\ref{PMeq} and it would be possible to increase the dissipative behavior of the shown parameters by adding more iterations to the relaxation filtering methodology proposed.

\begin{figure}[!t]
\centering
\mbox{
\subfigure[$N=512$]{\label{Figure7a}\includegraphics[width=0.5\textwidth,trim=4 4 6 6,clip]{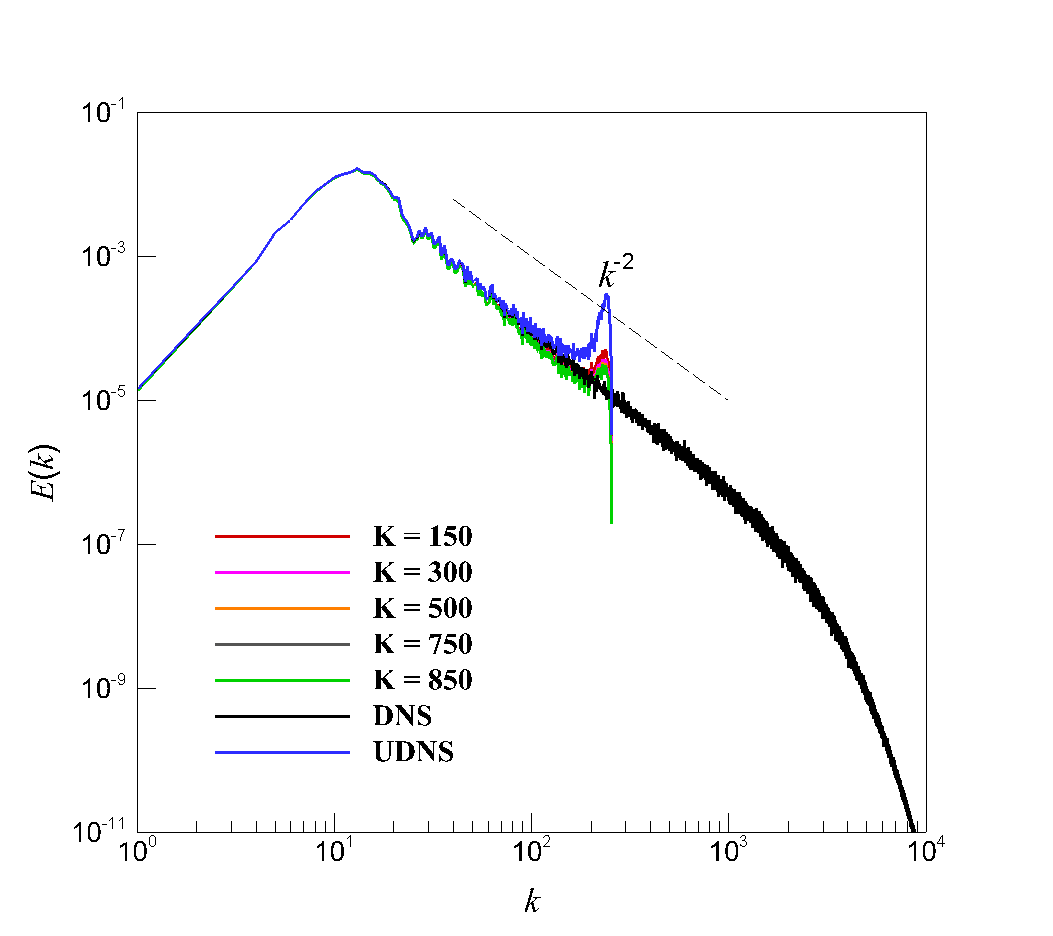}}
\subfigure[$N=1024$]{\includegraphics[width=0.5\textwidth,trim=4 4 6 6,clip]{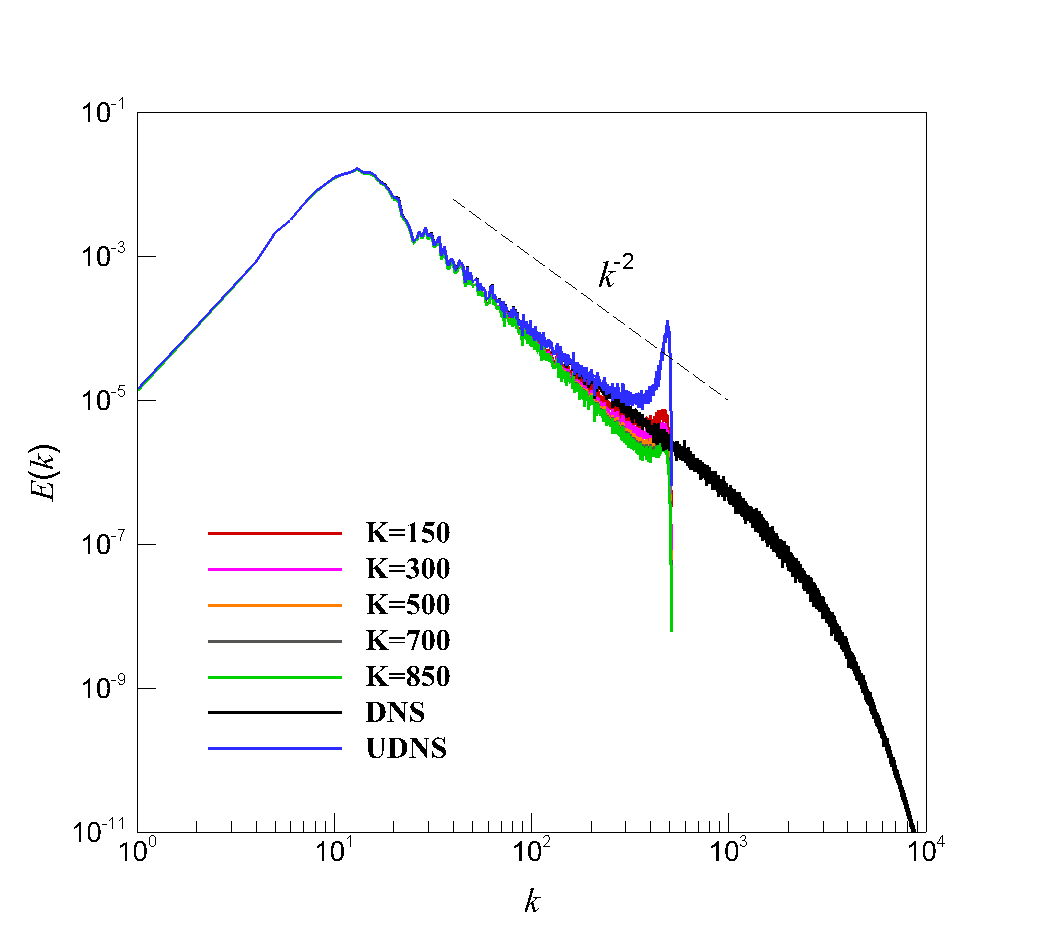}}}
\mbox{
\subfigure[$N=2048$]{\label{Figure7b}\includegraphics[width=0.5\textwidth,trim=4 4 6 6,clip]{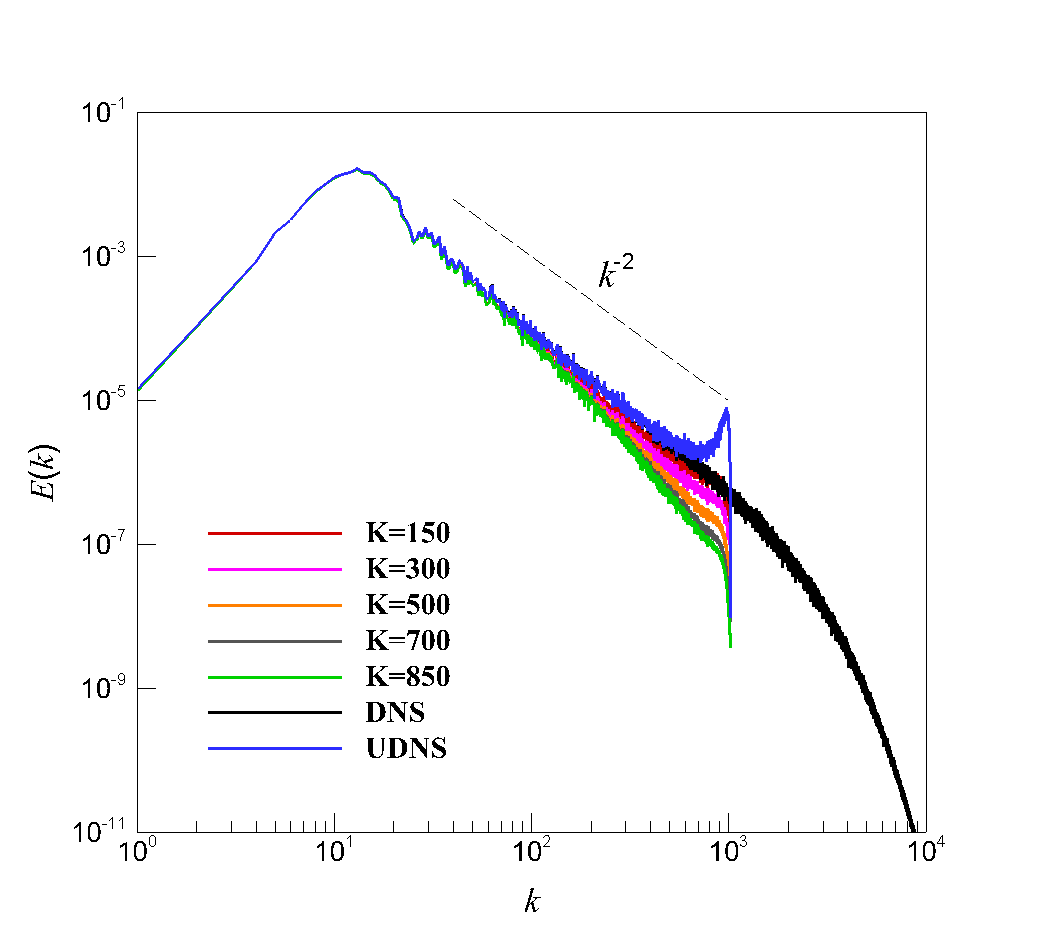}}
}
\caption{Effect of diffusion constant $K$ for diffusivity kernel 3.}
\label{Figure7}
\end{figure}

Fig.~\ref{Figure7} shows the effect of the diffusion constant for the kernel proposed by Eq.~(\ref{CondFunc2}). Similar trends are seen for both free parameters. It can be noted here that the parameter $\kappa$ which controls the ratio of pseudotime step to physical time step is optimal for the Burgers turbulence case at $\kappa = 0.0025$ across the different kernels examined in this work. The prescribed value of $\kappa$ ensures consistency in energy accumulation control with increasing resolution. A common theme that can be seen through the sensitivity analyses carried out for the different kernels is a good approximation of the inertial range scaling. In comparison, Fig.~\ref{Figure9} shows the results obtained using the selective relaxation filter given by Eq.~(\ref{bogey}), which diminish any grid-to-grid oscillations effectively but causes a degradation of the inertial range captured. The proposed framework identifies this aspect as a major potential area of improvement. This is possible due to a greater degree of freedom to specify the dissipation at different modes of the dispersion errors.

\begin{figure}[!t]
\centering
\mbox{
\subfigure[$N=512$]{\includegraphics[width=0.5\textwidth,trim=4 4 6 6,clip]{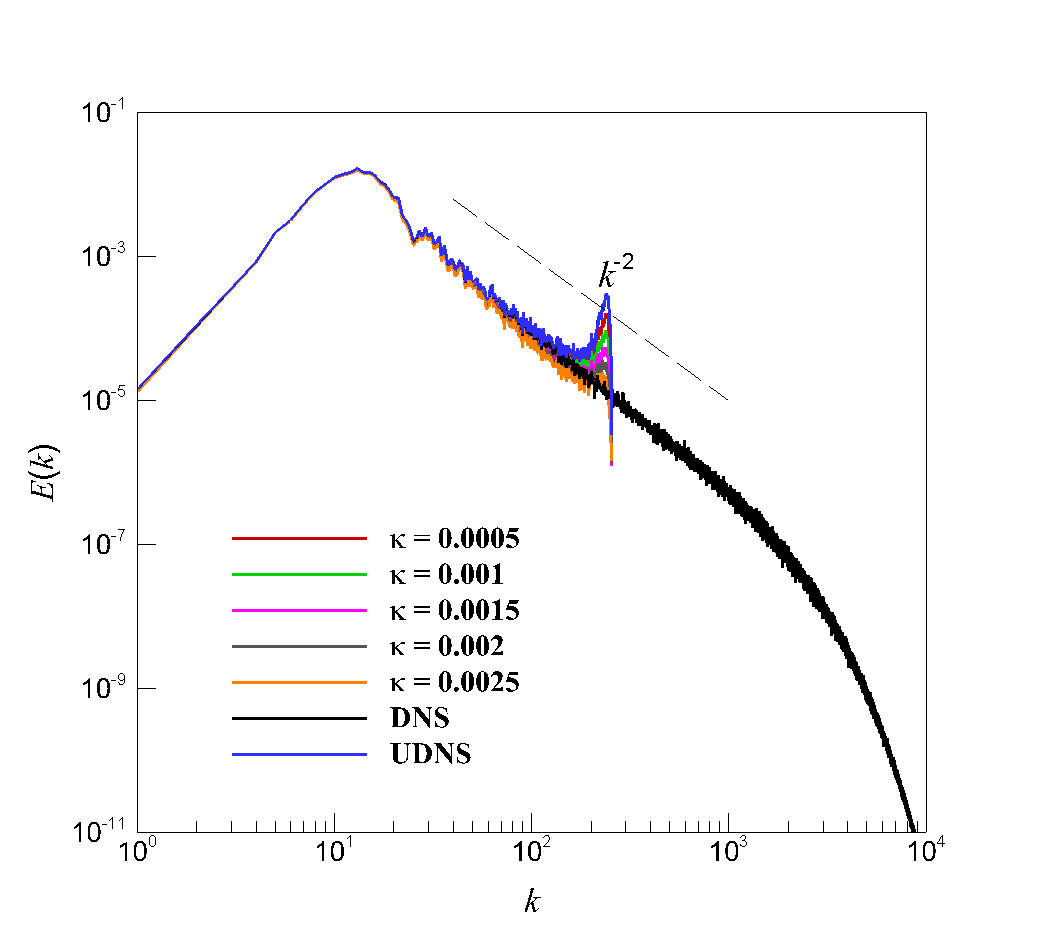}}
\subfigure[$N=1024$]{\includegraphics[width=0.5\textwidth,trim=4 4 6 6,clip]{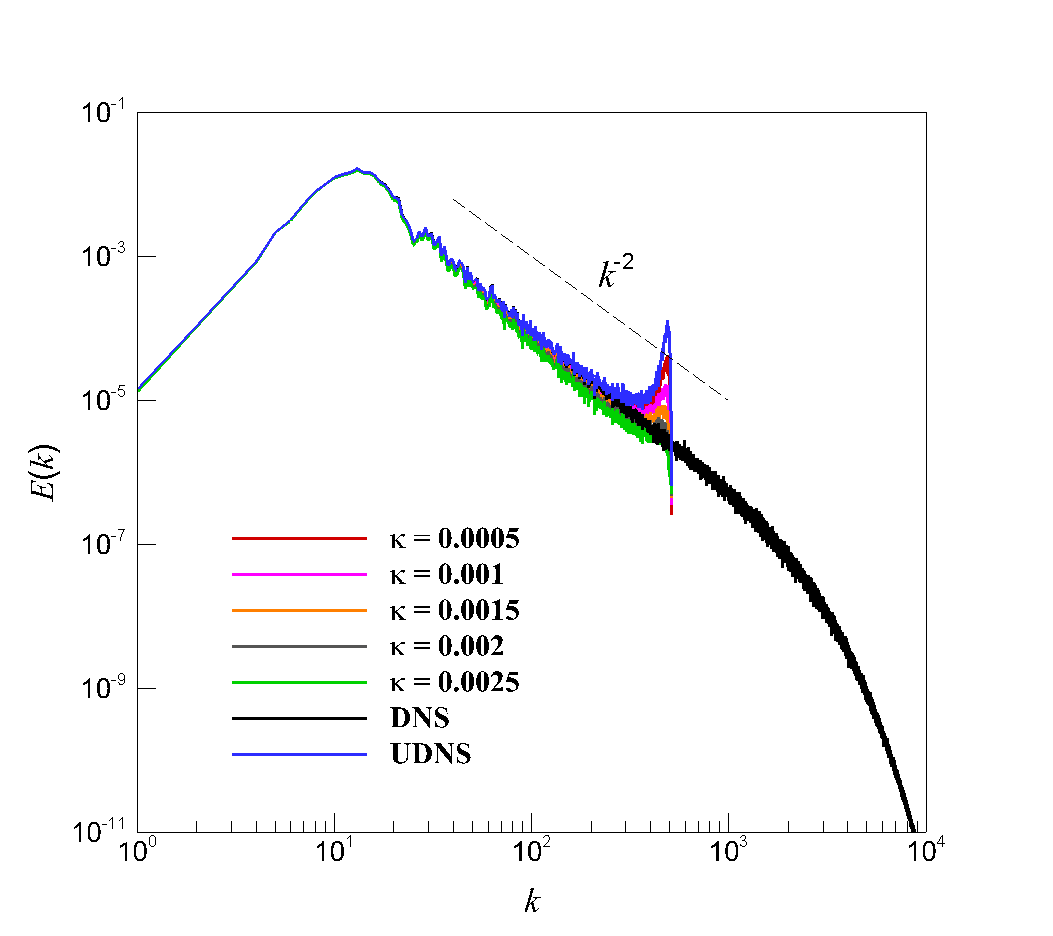}}}
\mbox{
\subfigure[$N=2048$]{\includegraphics[width=0.5\textwidth,trim=4 4 6 6,clip]{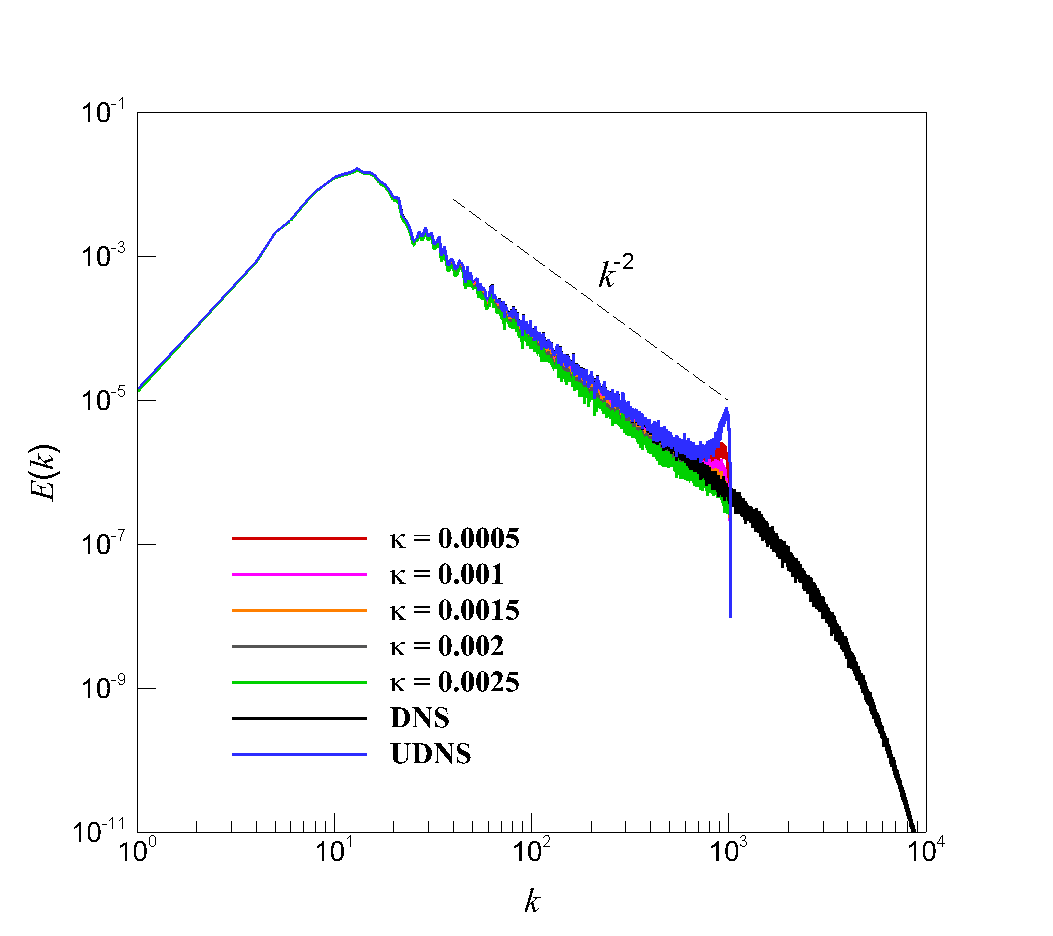}}
}
\caption{Effect of $\kappa$ for diffusivity kernel 3.}
\label{Figure8}
\end{figure}

\begin{figure}[!t]
\centering
\mbox{
\subfigure[$N=512$]{\includegraphics[width=0.5\textwidth,trim=4 4 6 6,clip]{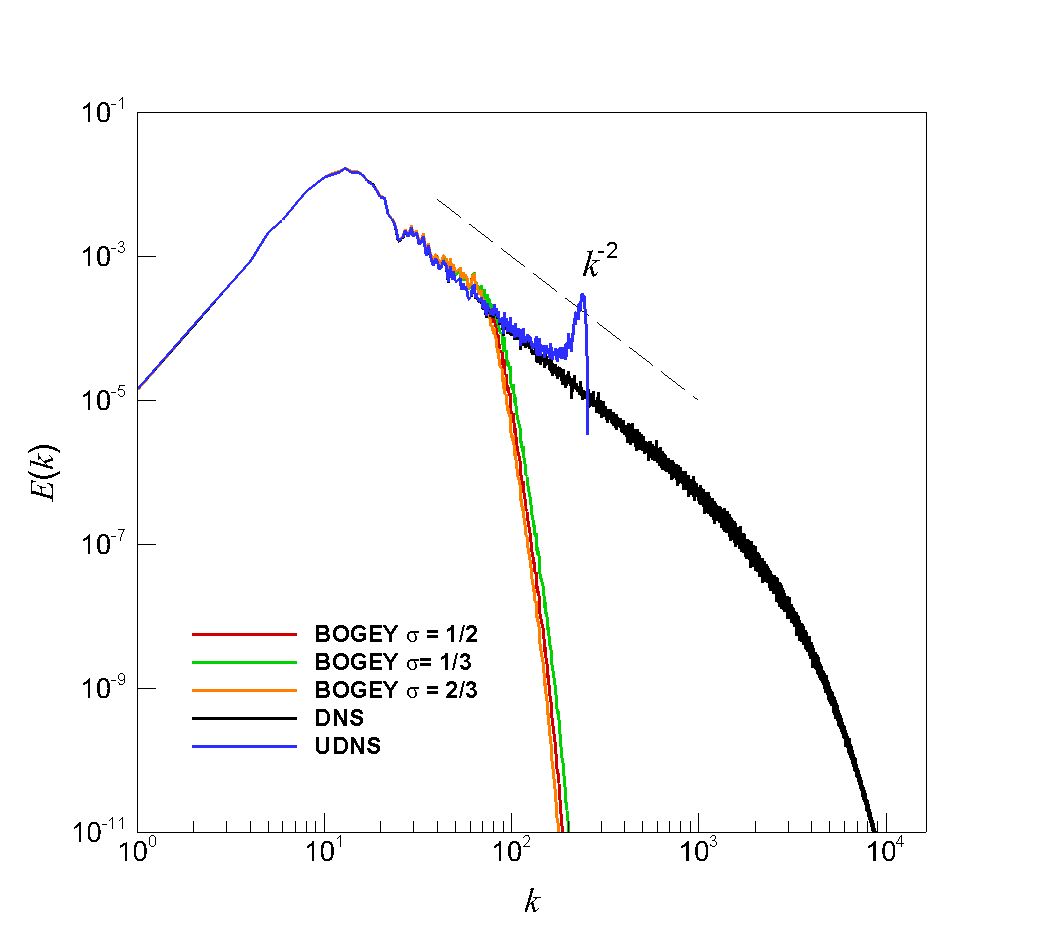}}
\subfigure[$N=1024$]{\includegraphics[width=0.5\textwidth,trim=4 4 6 6,clip]{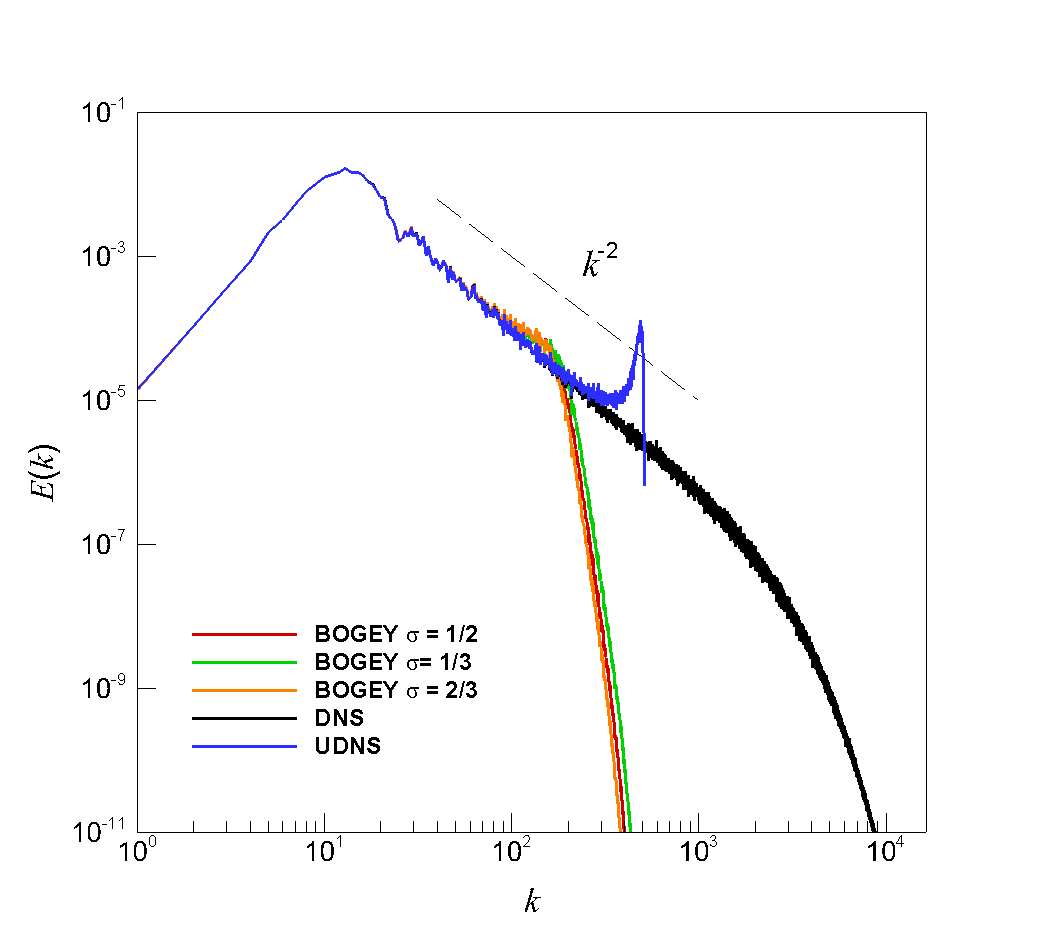}}}
\mbox{
\subfigure[$N=2048$]{\includegraphics[width=0.5\textwidth,trim=4 4 6 6,clip]{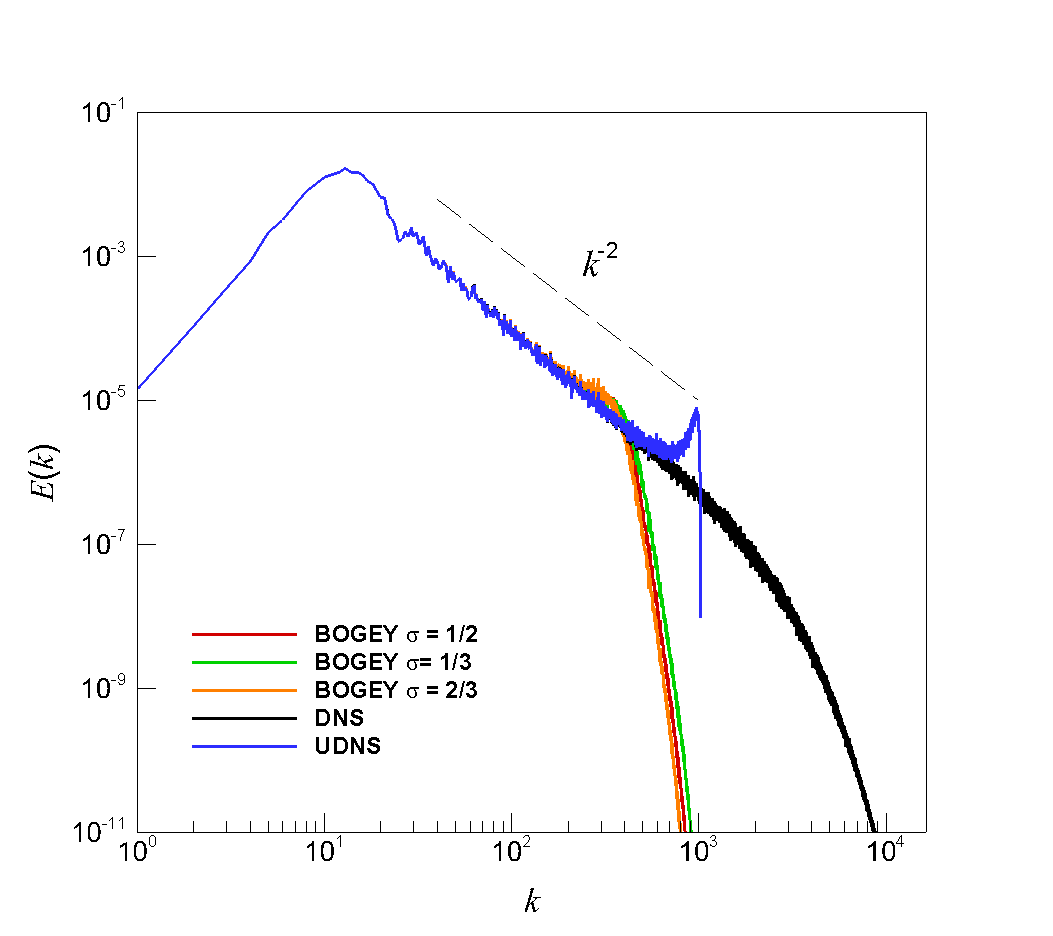}}
}
\caption{Performance of the selective relaxation filter given by Eq.~(\ref{bogey}).}
\label{Figure9}
\end{figure}



\section{Conclusion}


This study presents an explicit relaxation filtering and shock capturing framework based on the celebrated Perona-Malik anisotropic diffusion methodology for solving the Burgers turbulence problem. A detailed sensitivity analysis is presented to determine the dissipative action of the free modeling parameters of this framework. Results are compared against fully resolved DNS and UDNS computations to determine its performance as an effective closure model. The proposed model presented here uses a conductivity kernel that controls a dissipation imparted through the successive iterations of a parabolic partial differential equation analogous to the heat equation. The number of iterations is kept at a default value of one to control the dissipation of the model. This is because the pseudotime step used in the iteration of the parabolic equation needs to be sufficiently small to prevent destruction of the inertial range scaling. The shock capturing ability of the proposed framework is tested with a single mode sine wave as the initial condition of the governing non-linear viscous Burgers equation. The closure performance of the model is assessed solving the Burgers turbulence problem consisted of defining 32 sample fields specified by an initial energy spectrum. A sensitivity analysis is described to detail the effects of the control parameters $K$ and $\kappa$ in the model.


The shock capturing study indicates that the anisotropic diffusion characteristics of the explicit filter enables efficient dissipation of the grid to grid oscillations (an expected result due to the roots of this method in the field of edge detection in image processing). Comparing against a selective filtering approach that uses an optimized scheme on wide stencil, we show that a larger inertial range can be obtained by the proposed Perona-Malik relaxation model using a simple compact stencil scheme in an efficient way. The sensitivity analysis of the Burgers turbulence case indicates that the free modeling parameters are effective in control of the dissipation imparted by the filter and an increase in the values of both $K$ and $\kappa$ causes increased dissipative behavior. We found that the effect of $K$ has more pronounced due to the nature of the sharp gradients (shocks) in Burgers turbulence. We observe that certain combinations of the free modeling parameters may give us accurate inertial range of the energy spectrum even for the coarsest resolutions. However, an adjustment in the parameters would be necessary in different flows. This suggests that a dynamic procedure to characterize these modeling parameters would be an important open question, which will be addressed in our forthcoming research.


To summarize, we can conclude that the proposed closure proves an efficient methodology to remove aliasing error and capture shocks for the one-dimensional Burgers turbulence case due to the flexible dissipation characteristics imparted by the free modeling parameters. The use of a one-iteration relaxation filter at the end of each timestep also implies a minimal addition to the computational expense of the overall unresolved DNS computation and provides a minimal dissipation to eliminate higher frequency contents via the anisotropic dissipation process. As a further investigation in this direction, a dynamic modeling could be introduced for the free modeling parameters to add the requisite amount of dissipation for all resolutions. Although, the presentation of the proposed closure is given for the case of Burgers turbulence, its generalization to the 3D Navier-Stokes equation is straightforward. We believe that general conclusions from our study solving the Burgers equation will still hold since it retains some important properties of the Navier-Stokes equations.

\section*{References}

\bibliography{mybibfile}

\end{document}